\documentclass[preprint,12pt]{elsarticle}
\usepackage{graphicx}
\usepackage{multicol,multirow}
\usepackage{amsmath,amssymb,amsfonts}
\usepackage{mathrsfs}
\usepackage{amsthm}
\usepackage{rotating}
\usepackage{appendix}
\usepackage{ifpdf}
\usepackage[T1]{fontenc}
\usepackage{newtxtext}
\usepackage{newtxmath}
\usepackage{textcomp}
\usepackage{xcolor}
\usepackage{lipsum}
\usepackage[colorlinks,allcolors=blue]{hyperref}

 \usepackage{ulem}
 \usepackage{soul}
 \usepackage{cancel}

 \usepackage{subcaption}
\usepackage{tocloft}
\usepackage{geometry}
\theoremstyle{definition}

\numberwithin{equation}{section}

\newcommand*{\be}{\begin{equation}}
\newcommand*{\ee}{\end{equation}}
\newcommand*{\bea}{\begin{eqnarray}}
\newcommand*{\eea}{\end{eqnarray}}
\newcommand{\sn}{\text{sn}}
\newcommand{\cn}{\text{cn}}
\newcommand{\dn}{\text{dn}}

 \newcommand{\ts}{\thinspace}

\usepackage{amssymb}
\usepackage{amsmath}

\journal{Physica D}

\begin{document}

\begin{frontmatter}



\title{Swinging Waves in the Ablowitz-Ladik Equation}


\author[uct,jinr]{I.~V.~Barashenkov}
\author[uct]{Frank~S.~Smuts}

\affiliation[uct]{
  organization={Centre for Theoretical and Mathematical Physics, University of Cape Town},
  addressline={Rondebosch 7701},
  country={South Africa}
}

\affiliation[jinr]{
  organization={Joint Institute for Nuclear Research},
  addressline={Dubna 141980},
  country={Russia}
}

\begin{abstract}
We construct a novel family of exact cnoidal wave and soliton solutions of the focusing and defocusing Ablowitz-Ladik equations. 
Unlike cnoidal waves  that were obtained by earlier authors, the phase variable of the new
solutions exhibits a  nonlinear dependence on time and site number;  the wave ``swings". 
Our approach hinges on the existence of a two-point map governing the absolute value of the complex field; this map gives rise to standing waves centred arbitrarily relative to the lattice sites. 
Having derived stationary solutions, we use these as a basis for constructing waves with nonzero velocity.
The localised members of the new family comprise dark solitons with the nontrivial asymptotic behaviour.
We identify periodic and quasiperiodic patterns and establish an explicit quantisation rule for 
 the velocity of the wave circulating in a 
closed loop of $N$ sites.
\end{abstract}



\end{frontmatter}



\section{Introduction} 
 
The Ablowitz-Ladik equations \cite{AL1,AL2}, 
 \bea
 i  \frac{ \partial \psi_n}{\partial \tau} 
  + \frac{\psi_{n+1}- 2 \psi_n + \psi_{n-1}}{\mu^2}   + \sigma  (\psi_{n+1}+ \psi_{n-1}) |\psi_n|^2 =0,
  \label{B1}
 \eea
 represent the only known integrable semi-discretizations of the nonlinear Schr\"odinger equations, and are therefore of considerable
 mathematical interest   \cite{APT}. In \eqref{B1}, $\sigma$ is a sign factor, with $\sigma =1$ defining the focusing and $\sigma=-1$ defocusing equation. 
 The  Ablowitz-Ladik equations provide insights into the general properties of nonlinear lattices 
 \cite{HT,EJ,kevrekidis_discrete_2009}
 and serve as a starting point for perturbation expansions into nonintegrable systems
 \cite{cai1,cai2,melvin_discrete_2009,sullivan_kuznetsov-ma_2020,Mithun1, Hennig1,Hennig2, Mithun2}.
The present study aims to expand the class of their physically meaningful, mathematically tractable solutions. 
These describe periodic, quasiperiodic, and localised structures.

The inverse scattering problem for the Ablowitz-Ladik equation in
 the quasiperiodic case was first treated in Ref \cite{bogolyubov_discrete_1982} and  \cite{ahmad_quasi-periodic_1987}. 
 Subsequently, the authors of  Ref \cite{miller_finite_1995} used
  algebraic-geometric methods  to express quasiperiodic solutions in terms of the Riemann $\theta$-functions 
  with dependence on one or more travelling-wave combinations.
 The Fay identity was then leveraged to construct solutions for the entire Ablowitz-Ladik hierarchy \cite{vekslerchik_finite-genus_1999}. 
 (For solitons, see  \cite{AL1,AL2,KV,Chubykalo, APT,ABP, Prinari_solo,Prinari_Vitale}.)

The modulation instability of spatially extended states, such as  cnoidal waves
often triggers anomalous events, including blowups and rogue wave formation. 
There is an extensive literature on the anomalous solutions of the Ablowitz-Ladik equations \cite{Ankiewicz, Ohta1, Ohta2, Ortiz, ChenPeli_2, Santini, Prinari}.

In the present work, we construct solutions that depend on a single travelling-wave combination and are expressible 
in terms of the Jacobi elliptic functions and 
Legendre's elliptic integral.
 These solutions present crucial advantages over their equivalent representation in terms of the $\theta$-functions. They are straightforward to visualize numerically, easy to interpret physically and highly amenable to analyses of their linear and orbital stability. 

 The transformation
\be 
 \label{B2} 
 \psi_n(\tau)= \frac{U_n(t)}{\mu}               e^{-2it}, 
 \quad \tau= \mu^2  t
 \ee
 casts equation  \eqref{B1} into a nondimensional form
\be
i \frac{\partial U_n}{\partial t} + (U_{n-1} + U_{n+1}) (1 + \sigma |U_n|^2)=0.
\label{A1}
\ee
Elliptic-function solutions of equation \eqref{A1}, comprising an exponential carrier wave modulated by a real periodic amplitude, 
\be
U_n(t)= \phi (n-Vt) e^{i( \Omega t   + Qn)},
\label{B3}
\ee
were reported in Ref \cite{LaiZhang}.
(See also \cite{ChenPeli_1,ChenPeli_2}.) 
The list of  \cite{LaiZhang} was expanded to include sequences with $|U_n|^2=1$ \cite{HuangLiu}. Unlike the traveling waves of \cite{ChenPeli_1,ChenPeli_2,HuangLiu,LaiZhang}, the wave trains we present below feature a phase with nonlinear dependence on $t$ and 
$n$. In the terminology of \cite{bottman_elliptic_2011,ChenPeli_1,ChenPeli_2,deconinck_stability_2017} these are  "nontrivial phase solutions": intrinsically complex structures that cannot be separated into an exponential carrier wave and its real envelope \cite{carr_stationary_2000}. Our wave trains comprise all previously obtained traveling-wave solutions as particular cases.

The infinite-period  limit of our Jacobi-function solutions yields new classes of dark solitons of the defocusing Ablowitz-Ladik equation. 
Previously, the authors of \cite{ABP,KV,VK, Chubykalo,Prinari_solo,Prinari_Vitale} constructed a two-parameter family of travelling localised solutions
under the boundary conditions 
\be
\label{C2}
U_n \to \sqrt A e^{2i(1-A)t \pm  i \vartheta}   \quad \mbox{as} \quad   n \to \pm \infty,
\ee
 with constant  $\vartheta$. Our dark solitons have a more general asymptotic behaviour 
\be 
\label{C1}
U_n \to \sqrt{A}  e^{ i (\Omega t + Q  n)  \pm i  \vartheta}   \quad \mbox{as} \quad   n \to \pm \infty,
\ee
where the asymptotic frequency $\Omega= 2(1-A) \cos Q$, the amplitude of the background wave satisfies 
$0 < A<1$, and the wavenumber $Q$  and phase shift $\vartheta$ can be chosen arbitrarily.

The paper is organised into six sections. In section \ref{SW} we construct stationary (that is, nonpropagating) wave solutions. The crux of our approach is laid out in its first part
where we derive an integrable  difference equation for the absolute value of the Ablowitz-Ladik variable $u_n$. 
In the next section (section \ref{TW})  the analysis is extended to waves travelling with nonzero velocities. 
The primary result here is a closed-form expression for the phase of the wave. 
Solitons -- spatially localised solutions ---- are of special importance for applications; these are dealt with in section \ref{solitons}.
Another class of solutions of physical significance is selected by the requirement of periodicity. We consider stationary and travelling periodic waves in section \ref{periodic}.
Finally, section \ref{Conclusions} summarises conclusions of this study and suggests links to other nonlinear systems.

\section{Standing waves}
\label{SW} 

\subsection{Amplitude map}
\label{AM} 

Letting $U_n(t)= u_n e^{ 2i \omega t}$ with time-independent $u_n$  gives an equation for standing waves, that is,
stationary, nonpropagating, patterns:
\be
(u_{n-1} + u_{n+1}) (1 +  \sigma |u_n|^2)= 2 \omega u_n.
\label{A2}
\ee
Equation \eqref{A2} has two invariants  \cite{ChenPeli_2}:
\be
\label{A3}
I_1= \frac{i}{2}  (u_n u_{n+1}^*  - u_n^* u_{n+1} ) 
\ee
and
\be
I_2= |u_{n+1}|^2+|u_n|^2 + \sigma  \, |u_n u_{n+1}|^2      - \omega \,  (u_n u_{n+1}^*+ u_{n+1}  u_n^*).
\label{A4}
\ee
Writing  $u_n=\phi_n e^{i \theta_n}$, with $\phi_n= |u_n|$, 
the  invariants acquire the form
\be
I_1=   \phi_n \phi_{n+1} \sin (\theta_{n+1} - \theta_n)
\label{A5} 
\ee
and 
\bea
I_2= \rho_n+ \rho_{n+1} + \sigma  \rho_n \rho_{n+1} - 2 \omega \phi_n \phi_{n+1} 
 \cos (\theta_{n+1}-\theta_n),
\label{A500}
\eea
where $\rho_n = |u_n|^2$. 
Eliminating the phase variable between \eqref{A5} and \eqref{A500} gives
\be
(\rho_n+ \rho_{n+1} +   \sigma \, \rho_n \rho_{n+1} -I_2)^2=
4 \omega^2 (\rho_n \rho_{n+1} -I_1^2).
\label{A6}
\ee
Subtracting the left- and right-hand sides of \eqref{A6} from their expressions with $n \to n-1$,
we obtain a symmetric second-difference equation 
for the amplitude $\rho_n$:
  \be
 \label{E2}
 \rho_{n+1} - 2 \rho_n + \rho_{n-1} =   - \frac{4\sigma   \omega^2}{(\rho_n + \sigma)^2}  + 2 \frac{1   +\sigma I_2 + 2 \omega^2}{\rho_n +\sigma}  - 2(\rho_n  + \sigma).
 \ee 
The two-point map \eqref{A6} serves as a conservation law for the difference equation \eqref{E2}.

The existence of a two-point map indicates an effective ``translation invariance" of equation \eqref{E2}, meaning 
that there is a continuous function  $\rho(x)$ such that  the sequence $\rho_n=\rho(x_n)$
with $x_n=\mu n +  x^{(0)}$ solves  \eqref{E2} for any value of the translation parameter $x^{(0)}$. 
   (See \cite{Panos,BOP}.)  The function $\rho(x)$ should satisfy the advance-delay equation 
     \be
 \label{E222}
 \rho_+ - 2 \rho + \rho_- =   - \frac{4\sigma   \omega^2}{(\rho + \sigma)^2}  + 2 \frac{1   +\sigma I_2 + 2 \omega^2}{\rho +\sigma}  - 2(\rho  + \sigma),
 \ee 
 where $\rho_\pm$ is our short-hand notation for $\rho(x \pm \mu)$.

Equation \eqref{E222} can be solved explicitly. 
As a candidate solution 
 we try a class of periodic wavetrains
 \be 
 \label{E1}  
 \rho(x)=   A - B\ts \cn^2 (x, m), \quad A \geq 0.
 \ee
Here, the exact definition of the elliptic modulus $m$ ($0 \leq m \leq 1$)
is specified by the Jacobi delta identity:
 \[
 \dn^2 (x,m) + m \sn^2(x,m)=1.
 \]
 (In what follows, we routinely omit the dependence of the Jacobi functions on the  modulus; that is, we write $\sn y$ for $\sn(y,m)$, 
 $\cn y$ for $\cn (y,m)$, and so on.)
 The Ansatz \eqref{E1} is proposed based on the analogy with the cnoidal-wave solution of the differential nonlinear Schrödinger equation.
 (See \ref{cnoNLS}).

With the choice of \eqref{E1},  the left-hand side of equation \eqref{E2} can be represented as a rational function of $\rho$:
\be
\label{F3}
\rho_+ +\rho_- = 2 A + \frac{2B} {m^2 \sn^4 \mu}  \frac{P_2(\rho)}{(\rho-\rho_*)^2},  
\ee
where 
\[
P_2(\rho)=       B      \ts     \cn^2 \mu  \ts(\rho-A) + m \ts \sn^2 \mu  \ts \dn^2 \mu   \ts  (\rho -A+B)    \left(\rho -A+\frac{m'}{m} B  \right),
\]
with $m' = 1-m$,
and
\be
\label{F4}
\rho_*= A +B \frac{ \dn^2 \mu}{ m \sn^2 \mu}.
\ee

Comparing the positions of the pole in the left- and right-hand sides of \eqref{E222} we observe that $\rho_\star= -\sigma$. Equating the corresponding coefficients of the Laurent expansions then gives
\bea
B= - m \frac{\sn^2 \mu}{\dn^2 \mu}     (A+ \sigma), 
\label{F5}
\\
\label{F51}
\omega^2 =   \frac{\cn^2 \, \mu}{\dn^4 \mu}       (1+ \sigma A)^3
\eea
and 
\be
 I_2=  \frac{  \sigma \dn^2 \mu     -                   (2A + \sigma m \sn^2 \mu) \cn^2 \mu }{\dn^4 \mu} (1 + \sigma A)^2 - \sigma.
 \label{F6}
\ee
The value of the invariant $I_1^2$ can be determined from equation \eqref{A6}:
\be 
\label{F7}
I_1^2= \frac{A(A - B)}{\dn^2 \mu} \Big( m^\prime \sn^2 \mu - \sigma A  \cn^2 \mu\Big).
\ee
In \eqref{F5}-\eqref{F7}, the parameter $A$ is constrained by
 \bea 
 0   \leq  A   \leq   m^\prime  \frac{\sn^2 \mu}{\cn^2 \mu},   & \quad  \sigma=  &1;   \label{F105}   \\
 m \ts  \sn^2 \mu  \leq  A < 1, & \quad \sigma=-  & 1.
 \label{F106}
 \eea
It  is worth noting that $B$ is negative for $\sigma=1$ but positive for $\sigma=-1$.

Note  that equation  \eqref{F51}  determines only the absolute value of the frequency $\omega$  while its sign can be chosen arbitrarily.
 This is due to the maps \eqref{A6} and \eqref{E2} 
 being invariant under the change $\omega \to -\omega$.  The stationary solutions corresponding to frequencies 
 $\omega$ and $\tilde{\omega}= -\omega$ are related by the 
    staggering transformation 
    \be 
    \label{stag}
    u_n \to \tilde{u}_n = (-1)^n u_n.
    \ee
    
The sign of $I_1$ can also be chosen arbitrarily.  Note that there is no correlation between the signs of $I_1$ and $\omega$. 
If $u_n$ is a solution of the stationary equation \eqref{A2} with a given $\omega$,  then
the sequence $u_n^*$ is another solution of that equation, with the same $\omega$  but opposite value of $I_1$.

Finally, writing $\rho+\rho_+$ and $\rho \rho_+$ as quadratic polynomials in $\zeta^{-1}$, where
\be
\zeta=
\zeta(x)= 1- m\thinspace \sn^2 \lambda\thinspace  \sn^2(x+ \lambda)
\label{F120}
\ee
and
\be
\lambda=   \frac{\mu}{2}, 
\label{alpha}
\ee
 we obtain
 \begin{subequations}
 \label{F108}
  \be
 \rho + \rho_+ +\sigma \rho \rho_+  - I_2= 
 -  4 \sigma (1+ \sigma  A)^2 
  \frac{      \cn \mu     }{ \dn^3 \mu }    \times (\cn \mu+ \dn \mu) \ts  \mathcal R(x),
 \label{F108a}
 \ee  
 where
 \be
 \mathcal R(x)=  \frac{1}{  \zeta(x)}
- \frac{\sigma A    \ts \cn \mu  \ts  \sn^2 \lambda+  \sn^2 \mu\dn^2\lambda}
{  \sn^2 \mu \ts \dn \mu \ts  (1-m \ts \sn^4 \lambda) }.
\label{G5}
\ee
\end{subequations}
We will use the representation \eqref{F108} in sections \ref{sec:constant_phase} and \ref{Phase}  below.

\subsection{Standing waves with constant phase
($I_1=0$) }
\label{sec:constant_phase}

Before constructing waveforms with variable phase,
we acknowledge three solutions reducible to real sequences by a global phase shift.
(In the rest of this subsection, we consider $u_n$ to be real.)
Each of these  solutions  has its phase increment 
$\theta_{n+1}-\theta_n$  consisting of any sequence of zeros and $\pi$'s. The corresponding invariant $I_1$ has to be equal to zero.

In the defocusing case  ($\sigma=-1$), the constraint $I_1=0$ requires
 $A= m \sn^2 \mu$. 
 Pairs of the adjacent terms with opposite signs 
  ($u_n u_{n+1}<0$)
can be identified using the relation 
\be
u_n u_{n+1}=
\phi_n \phi_{n+1} \cos (\theta_{n+1}- \theta_n) =  m \ts \sn^2 \mu
 \frac{\cn \mu  \ts\dn \mu} {\omega}  \times
\frac{ \sn^2(x_n+ \lambda) - \sn^2 \lambda}{\zeta(x_n)},
\label{G600}
\ee
which follows from equations \eqref{A500} and \eqref{F108}. Using a readily verifiable identity
\[
\sn^2(x+ \lambda) - \sn^2 \lambda= \zeta(x)  \ts \sn(x+\mu) \ts \sn \ts x,
\]
the product \eqref{G600} is transformed into
\be
u_n u_{n+1}=
  m \ts \sn^2 \mu
 \frac{\cn \mu  \ts\dn \mu} {\omega} 
 \sn(x_n+\mu) \ts  \sn  \ts  x_n.
\label{G60}
\ee

   Letting $\sigma=-1$ and $A= m \sn^2 \mu$
  in  the equation \eqref{F51}, we 
  consider its root $\omega= \cn \mu  \ts \dn \mu$.
    With this choice of $\omega$, the relation \eqref{G60} implies 
    that 
the inequality 
$u_n u_{n+1} <0$ is satisfied whenever $\sn (x_n) \sn (x_n+\mu)<0$ holds true. 
This observation, together with the expression \eqref{E1} for $|u_n|^2$,
 gives rise to a real solution of the defocussing stationary Ablowitz-Ladik equation \eqref{A2}:
\be
u_n=   \sqrt m \ts \sn(\mu,m) \sn(x_n, m),   \quad \omega= \cn(\mu, m) \dn (\mu,m).
\label{F10}
\ee

The staggering transformation \eqref{stag} then generates
a solution with an opposite value of $\omega$.

In the self-focussing situation $\sigma=1$, the zero invariant $I_1$ is achieved by
choosing  either $A=0$ or $A=m^\prime \sn^2 \mu / \cn^2 \mu$. In the latter case, 
equations \eqref{A500} and \eqref{F108} give
\be
u_n u_{n+1}= \frac{\dn \mu  \ts\sn^2 \mu}{\omega \ts \cn^4 \mu}      \times  \frac{m^\prime  + m \ts  \cn^2 \lambda  \ts   \cn^2 (x_n+ \lambda)}
 { \zeta(x_n)}.
\label{H1} 
\ee
Using an identity
\[
m^\prime + m  \ts\cn^2 \lambda   \ts \cn^2    (x+\lambda)=
\zeta(x)\ts \dn (x+\mu) \ts \dn \ts x,
\]
this becomes
\be
u_n u_{n+1}=  \frac{\dn  \ts \mu  \ts\sn^2 \mu}{\omega \ts \cn^4 \mu} \, 
\dn \ts x_n  \ts   \dn (x_n +\mu).
\label{H1} 
\ee
Choosing the positive $\omega$, $\omega= \dn \mu / \cn^2 \mu$, we conclude that $u_n u_{n+1}>0$.
Using equation \eqref{E1} we arrive at the real solution of the focussing Ablowitz-Ladik equation:
\be 
u_n= \frac{ \sn (\mu, m)}{\cn(\mu, m)} \dn(x_n, m), \quad
 \omega=  \frac{\dn (\mu, m)}{ \cn^2(\mu, m)}.
\label{F9}
\ee

In the former case  ($A=0$), we have 
\be 
\label{H5}
u_n u_{n+1}= \frac{m  \ts  \sn^2 \mu \ts \cn \mu}{\omega \ts \dn^4 \mu } \times \frac{\cn^2 \lambda - \dn^2 \lambda \ts \sn^2  (x_n + \lambda)}{\zeta (x_n)             }. 
\ee
Choosing $\omega= \cn \mu / \dn^2 \mu$,  
using the expression for $\rho_n=\rho(x_n)$ in \eqref{E1} and invoking 
 the identity 
 \[
 \cn^2 \lambda - \dn^2 \lambda \ts \sn^2  (x+\lambda)  = {{\zeta} (x)}  \ts  \cn (x+ \mu)   \ts  \cn \ts x,
 \]
  we arrive at  the standing wave
\be
u_n = \sqrt m  \ts \frac { \sn( \mu, m)}{\dn(\mu, m)}   \ts  \cn (x_n,m) , \quad   \omega=  \frac{\cn(\mu,m)}{\dn^2 (\mu, m)}.
\label{F8} 
\ee

Applying the staggering transformation \eqref{stag}  to the high- and low-frequency standing waves,
equations   \eqref{F9} and \eqref{F8}, we 
obtain solutions with an opposite value of $\omega$.

 The constant-phase solutions
  \eqref{F10},  \eqref{F9}, and  \eqref{F8} are well known in the literature on the Ablowitz-Ladik equations 
  \cite{LaiZhang, ChenPeli_2,      HuangLiu}.

\subsection{Varied phase  ($I_1 \neq 0$) }
\label{Varied_Phase}

Turning to nonzero values of $I_1$, we 
let $A$ belong to the open interval $\left( 0, m^\prime  \frac{\sn^2 \mu}{\cn^2 \mu} \right) $   
in the case of the  focusing  equation, and to  $\left( m \sn^2 \mu,  1 \right)$ in the defocusing case.
 In either of these situations, equation \eqref{F7} gives  $I_1 \neq 0$.
 The amplitude $\rho_n$ has to  remain nonzero for all $n$
  ---- for the vanishing of $\rho_n$ even for a single value of $n$ would imply $I_1=0$. 

  Any sequence $\{ \rho_n \}$  
 solving the amplitude equation \eqref{E2} 
 and satisfying $\rho_n \neq 0$ for all    $n=0, \pm 1, \pm 2, ...$,
 defines a sequence of
  phase increments $\{ \Delta_n  \}$
  by means of equations \eqref{A5} and \eqref{A500}.
  Here
  \be
  \Delta_{n+1} :=    \theta_{n+1}- \theta_{n}. 
\label{A701}
\ee 
Note that since  the increment 
 $\Delta_{n+1}$ is expessible as a continuous function  of $\rho_n$ and $\rho_{n+1}$,  where
 $\rho_n$ is given by $\rho(\mu n + x^{(0)})$,
 there is a continuous function $\Delta(x)$ such that $\Delta_{n+1}=\Delta(\mu n  + x^{(0)})$.

Assuming that  $I_1 >0$, equation \eqref{A5} implies that $\sin \Delta_{n+1}>0$. Without loss of generality, 
 $ \Delta_{n+1}$ can be chosen to lie in the interval $(0, \pi)$.
 Equations  \eqref{A5}, \eqref{A500} and \eqref{F108}
   provide then an explicit  expression for the increment:
   \bea
   \Delta_{n+1} = \mathrm{arccot}   \left[  \frac{\rho_{n+1}+\rho_n+\sigma\rho_{n+1}\rho_{n}-I_2}    {2\omega   I_1}     \right]    \phantom{,}    \nonumber
 \\
  =  \mathrm{arccot} \ts \left[
  2 (-\sigma) \frac{ (1+ \sigma A)^2}{\omega I_1}  \times \frac{\cn \mu (\cn \mu + \dn \mu)} {\dn^3 \mu} 
    \times \mathcal R(x_n)  
  \right],
\label{A703}  
\eea
where $\mathcal R(x)$ is as in \eqref{G5}.  
In the case where $I_1<0$, the increment is given by 
  \be  
  \Delta_{n+1} =  \mathrm{arccot} \ts \left[
  2 (-\sigma) \frac{ (1+ \sigma A)^2}{\omega I_1}  \times \frac{\cn \mu (\cn \mu + \dn \mu)} {\dn^3 \mu} 
    \times \mathcal R(x_n)
  \right]-\pi.
\label{A7030}  
\ee

  Once we have obtained the increments, the phase $\theta_n$ 
can be recovered by summation:
\[
\theta_n= \left \{ \begin{array}{rl}  \displaystyle{ \theta_0+ \sum_{m=1}^n } \Delta_{m}, &  n>0; \\ & \\
   \displaystyle{\theta_0- \sum_{m=n+1}^{0} } \Delta_{m}, & n<0,
\end{array} \right.
\]
where $\theta_0$ can be chosen arbitrarily. 
In the next section we will show that $\theta_n$ can be written as $\theta(x_n)$ with some continuous function $\theta(x)$. 
We will obtain a 
 close-form expression for $\theta(x)$ by solving 
 a differential  (rather than a difference) equation.

\begin{figure*}[t]
    \centering
    \begin{subfigure}[b]{0.49\textwidth}
        \centering
           \includegraphics[width=\textwidth]{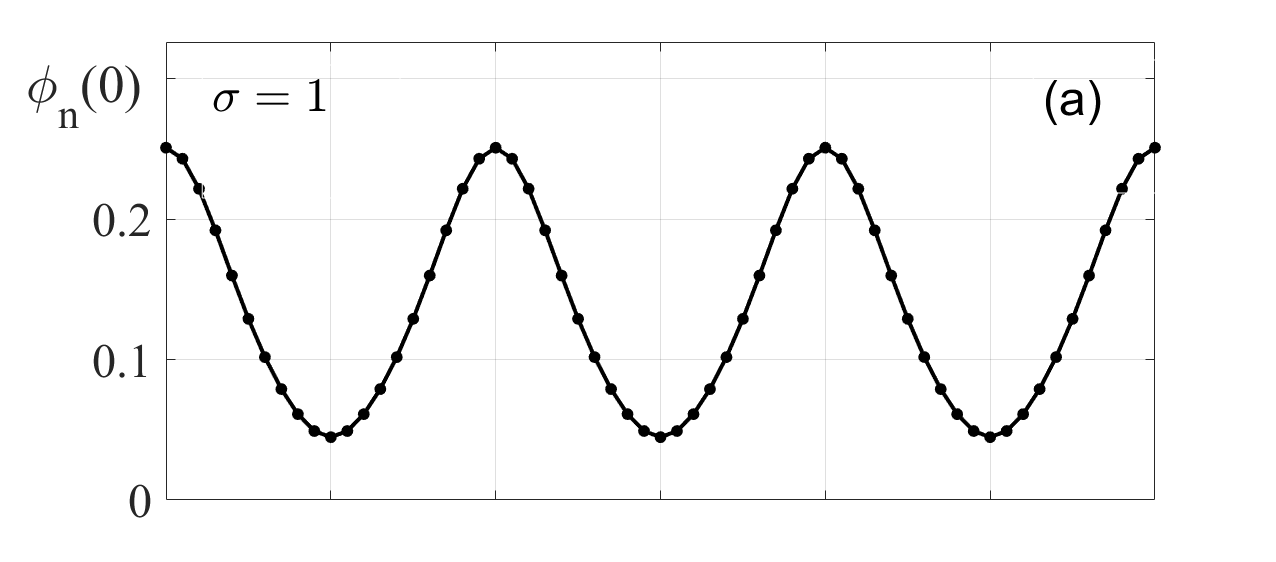}
               \end{subfigure}
    \begin{subfigure}[b]{0.49\textwidth}
        \centering
                \includegraphics[width=\textwidth]{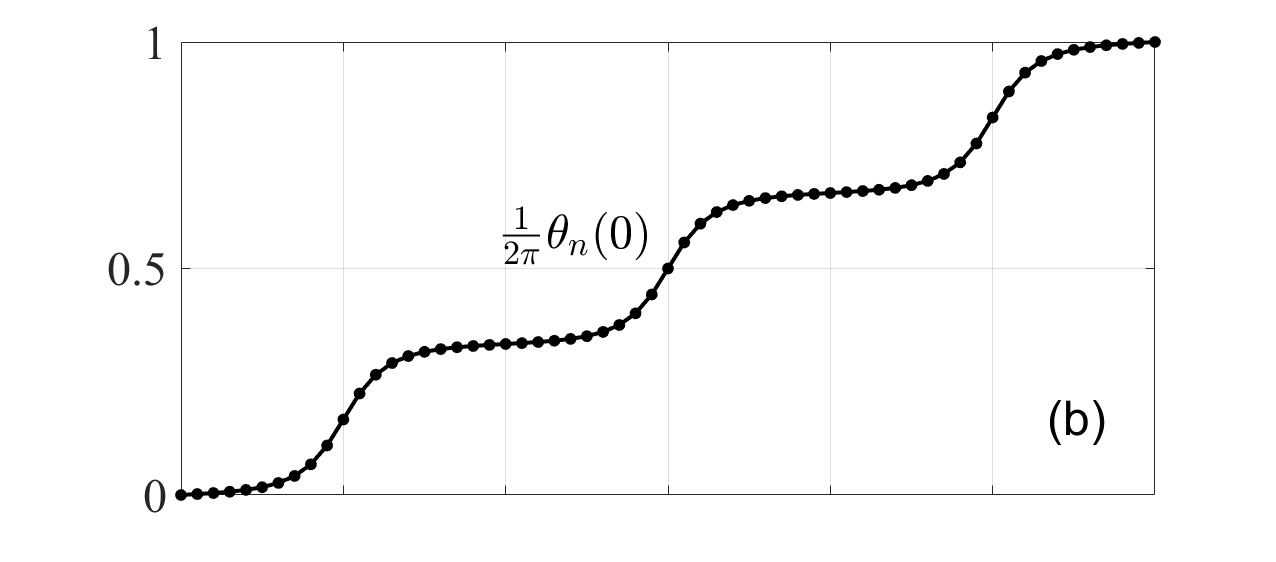}
                    \end{subfigure}\\
    \begin{subfigure}[b]{0.49\textwidth}
        \centering
                \includegraphics[width=\textwidth]{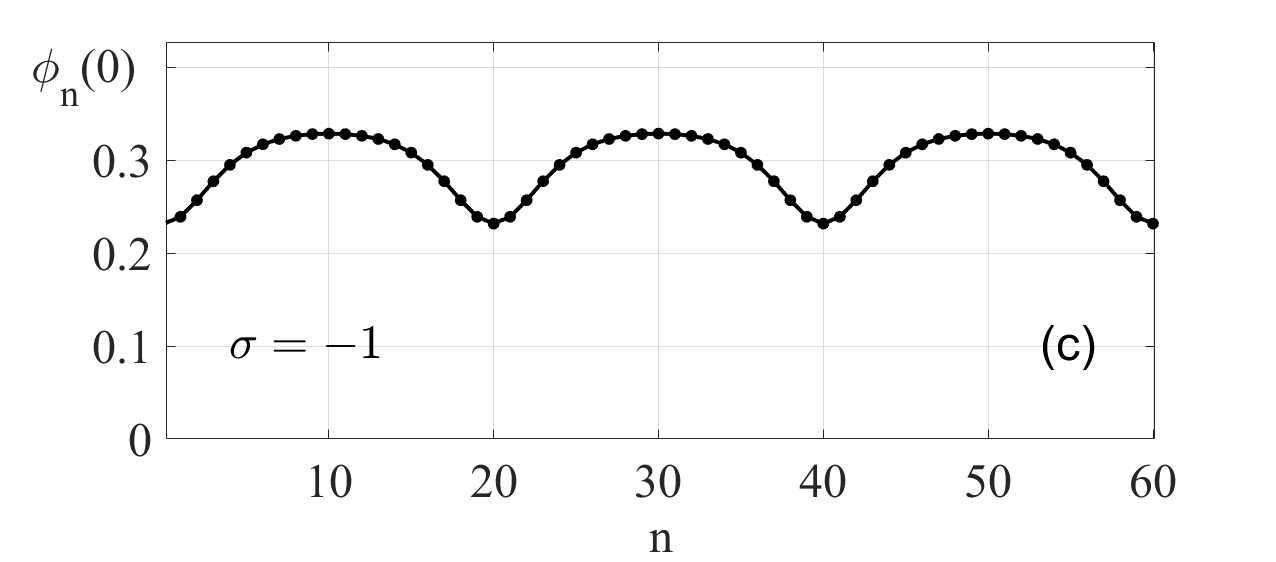}
        
    \end{subfigure}
    ~ 
    \begin{subfigure}[b]{0.49\textwidth}
        \centering
                \includegraphics[width=\textwidth]{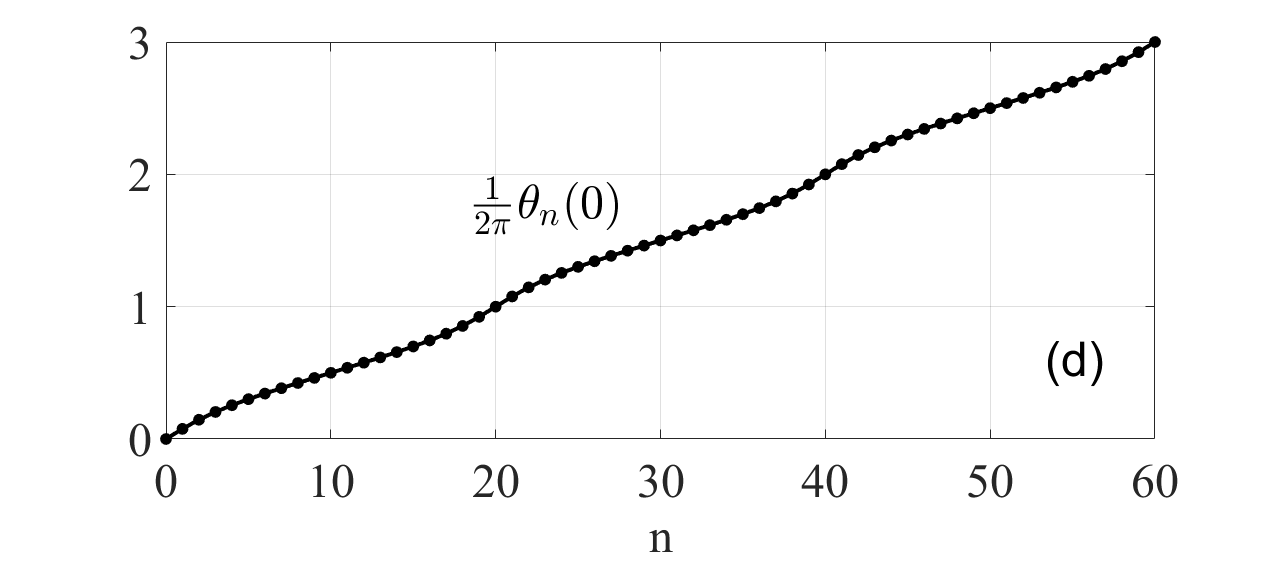}
    \end{subfigure}
    
    \caption{
    The modulus (a,c) and nonlinear phase (b,d) of periodic
 solutions to the stationary Ablowitz-Ladik equation \eqref{A2} with $\sigma=1$ (top) 
    and $\sigma=-1$ (bottom). Parameter values: $N=60,\;m=0.9,\;\mu=6K(m)/N$. The background amplitude $A$ is chosen to satisfy the periodicity condition
     $\theta_N-\theta_0=2\pi$
    for $\sigma=1$ and  $\theta_N-\theta_0=6 \pi$  for $\sigma=-1$.    
     (Equation \eqref{Y53} of section \ref{stanqua} gives, approximately,
      $A=A_1= 2.003\times 10^{-3}  $         
     for $\sigma=1$ and $A=A_3=1.080\times 10^{-1}$
      for $\sigma=-1$.
      These amplitude values occur as the horizontal coordinates of the blue blobs in Fig \ref{Aw} (a) and (b), respectively.) 
    }
    \label{fig:stat_foc_defoc}
\end{figure*}

 Fig \ref{fig:stat_foc_defoc} exemplifies the standing waves  $u_n=\phi_n  e^{i  \theta_n}$  by stationary
 solutions on a ring. (See section \ref{sec:numerical_implementation} for the implementation details.)

The stationary nonlinear-phase solution of the focussing or defocussing Ablowitz-Ladik equation is characterised by
4 parameters: $\mu$, $x_0$, $m$ and $A$. Of these, the intersite spacing $\mu$ is allowed to take any positive value.  
Once $\mu$ has been fixed, 
the centering parameter $x_0$ can be selected from the interval $(0, \mu)$. 
 
In view of the periodicity of the amplitude function \eqref{E1},
 with the period $2K(m)$, we can always assume that $0<\mu< 2K(m)$. For each given $\mu$, 
 this inequality establishes the range of admissible values of the elliptic modulus,  $m$.
  In the case of small spacings, $\mu< \pi$, the modulus is free to take any value between 0
 and 1. 
In the limit of $m \to 0$, choosing the root $\omega= \cos \mu (1+ \sigma A)^{3/2}$ in equation \eqref{F51} 
and letting, for definiteness, 
$I_1>0$,
 the standing wave reduces to an exponential:
\begin{equation}
u_n=\sqrt{A}\exp\left\{       in\arccos\left(\cos\mu  \sqrt{1+\sigma A}\right)   \right\}.
\end{equation}

For larger $\mu$, $\mu>\pi$, the interval of admissible elliptic moduli is shorter: $m_0 \leq m \leq 1$, where $m_0$ is the root of the equation $2K(m_0)=\mu$. 
Taking the limit $m\to m_0+0$ yields 
$u_n=0$ in the focusing case, and 
\begin{equation}
u_n=\sqrt{A}\exp\left\{   in\arccos\sqrt{1- A}\right\}          \label{Z12}
\end{equation}
in the defocusing situation. (In \eqref{Z12}, we selected the root $\omega= (1-A)^{3/2}$ of equation \eqref{F51} 
  and assumed $I_1>0$, for definiteness.)

\begin{figure*}[t]
    \centering
    \includegraphics[width=\textwidth]{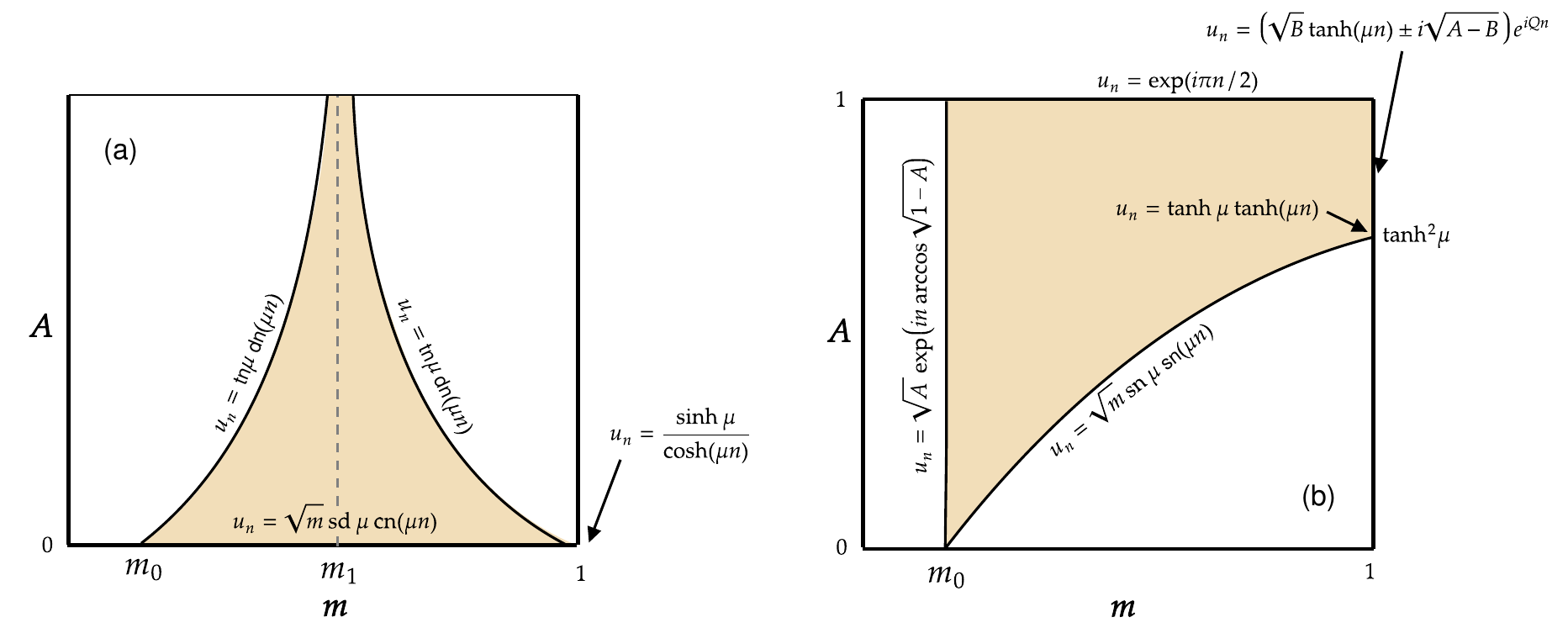}
    \caption{
    Parameter domain of the standing wave solutions on the $(m,A)$ plane, with the fixed $\mu$ and $x_0$, 
     in the focusing  (a) and defocusing (b) case.     In both cases we chose $\mu> \pi$ and let $x_0=0$.
  (The diagrams with $\mu< \pi$ look similar, the only difference being that $m_0=0$.)    
     In the focusing situation (a), the upper boundary is given by
         $A=(1-m)\frac{\sn^2(\mu,m)}{\cn^2(\mu,m)}$. In the defocusing case (b), the domain is bounded from below by $A=m\sn^2(\mu,m)$.
         Solutions indicated along the boundaries of the parameter domain are either real standing waves documented in section \ref{sec:constant_phase}
         (equations \eqref{F10}, \eqref{F9}, \eqref{F8})
         or linear-phase solutions with a constant modulus. The family of stationary dark solitons found along the $m=1$ boundary of the tinted domain in (b)
         is obtained in section \ref{solitons}.
                The notation $\mathrm{tn} (\mu,m)$ stands for $\sn (\mu,m) / \cn (\mu,m)$ and  $\mathrm{sd} (\mu,m)$ denotes
         $\sn (\mu,m) / \dn (\mu,m)$.  
        }
    \label{fig:phase_diagram}    
\end{figure*}

According to  \eqref{F105} and \eqref{F106}, 
the range of the amplitude $A$ of the standing wave depends on the sign of the 
nonlinearity in the Ablowitz-Ladik equation. 
In the focusing equation, the boundary of the parameter domain forms a narrowing cone extending 
to infinity  as  $m \to m_1$ (Fig \ref{fig:phase_diagram} (a)).
Here $m_1$ is defined as the root of  $K(m_1)=\mu$.
In the defocusing case, by contrast, the permissible pairs of $A$ and $m$ lie within a compact volume (Fig \ref{fig:phase_diagram} (b)).

\section{Traveling waves} 
\label{TW} 

 \subsection{Sliding wave: amplitude and velocity  } 
 \label{Mod}

Substituting the Ansatz 
\be
\label{A21}
U_n (t) = \phi(\xi_n)  e^{i \Theta_n}, \quad  
\Theta_n= \theta(\xi_n)+ kn+ (2 \omega+\nu)t
\ee
with 
 \be
 \xi_n= \mu (n-Vt)+ x^{(0)}
 \label{xin}
 \ee
  in the Ablowitz-Ladik equation \eqref{A1}  turns it into a pair of differential
  advance-delay equations
   \bea
    \mu V \theta^\prime \phi  -(2 \omega +\nu) \phi   +(1   + \sigma \rho)  [ \phi_-\cos(\theta_-- k)  
 + \phi_+ \cos(\theta_+ + k) ]=0,
 \label{A230}  \\
    \mu V \phi^\prime - (1  + \sigma  \rho) [ \phi_-  \sin(\theta_--k) + \phi_+ \sin(\theta_++ k)  ]=0.
\label{A23}  
 \eea  
 Here $\phi=\phi(\xi)$, $\theta= \theta(\xi)$, and
 $\rho(\xi) := \phi^2(\xi)$. 
 The plus and minus subscripts on $\phi$ and $\rho$ indicate the advance and delay of the argument, respectively:
  $\phi_\pm (\xi) := \phi(\xi \pm \mu)$, 
  $\rho_\pm (\xi) := \rho(\xi \pm \mu)$. By contrast, the same subscripts on $\theta$ denote an {\it increment\/} --- the finite difference incurred by
  advancing and delaying the argument:
  $\theta_\pm (\xi) := \theta(\xi \pm \mu)- \theta(\xi)$. 
The prime indicates differentiation with respect to $\xi$.

The system \eqref{A230}-\eqref{A23} with $V=k=\nu=0$ comprises 
the real and  imaginary part of 
the stationary  equation \eqref{A2} written in 
 the advance-delay form:
 \begin{align} 
  \phi_- \cos \theta_- + \phi_+  \cos \theta_+ = 2 \omega \frac{\phi}{1 + \sigma \rho},
\label{A501}  \\
 \phi_-  \sin   \theta_-    +   \phi_+ \sin \theta_+ =0.
\label{A502}   
\end{align}  
A family of solutions $\phi=\phi(\xi)$, $\theta_\pm= \theta_\pm(\xi)$
 to the stationary equations \eqref{A501}-\eqref{A502} were found in the previous section. 
Namely,  $\phi(\xi)=\sqrt{\rho(\xi)}$  is given by the periodic function \eqref{E1},
with  $B$ as in \eqref{F5}. The stationary increment
$\theta_+(\xi)$ was 
 denoted $\Delta(\xi)$ and $\theta_-(\xi)$ is expressible as     $ - \Delta(\xi-\mu)  $,
 with $\Delta(\xi)$  defined by equations \eqref{A5} and \eqref{A500}:
\begin{align}
 \sin      \Delta (\xi )= \frac{I_1}{\phi(\xi) \phi_+(\xi)},   \label{Y1} \\
 \cos \Delta(\xi)= \frac{ \rho(\xi) + \rho_+(\xi) + \sigma \rho(\xi) \rho_+(\xi) -I_2}
{ 2 \omega \phi(\xi) \phi_+(\xi)}.
\label{Y2}
\end{align}
Here $I_2$ is as in \eqref{F6}, while
$\omega$ and $I_1$ are given by one set of their values in \eqref{F51} and \eqref{F7}.   

The aim of this section is to
show that the phase $\theta(\xi)$ whose increment is given by \eqref{Y1}-\eqref{Y2}, 
together with the stationary amplitude \eqref{E1}, satisfy  the system \eqref{A230}-\eqref{A23} with properly chosen parameters  $k$, $V$, and $\nu$.
In essence, the standing wave can be propelled along the lattice without altering its form and exciting radiations.
An  explicit formula for   $\theta(\xi)$ will emerge as a  by-product of this analysis.

Assuming that $\phi$ and $\theta_\pm$ satisfy \eqref{A501}-\eqref{A502},
 we 
 use  \eqref{A502} to  simplify  equation \eqref{A23} 
 to
\be 
\mu V \phi' = \sin k (1 + \sigma \rho) 
( \phi_+  \cos \theta_+    
- \phi_-  \cos \theta_-   ),
\label{A231}
\ee
which can be further simplified 
 by noting the  relation 
\be
(\rho_+   -\rho_- ) (1  + \sigma \rho) = 2 \omega \phi
( \phi_+  \cos   \theta_+    - \phi_- 
\cos \theta_- ).
\label{A503}
\ee
The above relation results by substituting $n \to n-1$ in  \eqref{A500} and subtracting the original version of that equation.

Once the  phase variable has been eliminated with the help of \eqref{A503},  equation \eqref{A231} becomes
\be
 \omega   \mu V  \rho^\prime=  \sin k  \,  (\rho_+-\rho_-)(1 + \sigma  \rho)^2.
\label{A24}
\ee
By direct substitution, we verify that the function $\rho(\xi)$, defined by equation  \eqref{E1} and \eqref{F5} (where one just needs to replace $x$ with $\xi$)
satisfies 
equation \eqref{A24}  with
\be
V= 2 \frac{(1+\sigma A)^2}{\omega}  \,   \frac{\sn (\mu,m) \ts  \cn (\mu,m)}{\mu \ts \dn^3(\mu,m)} \,  \sin k.
\label{A26} 
\ee
Here $\omega$ is as in \eqref{F51}.

 \subsection{Phase}
 \label{Phase}
 
Equation  \eqref{A230} introduces a new variable that was absent in the stationary equations: the phase $\theta(\xi)$.
Note that  the phase  cannot  be uniquely reconstructed from its increment.
(Indeed, the functions $\theta(\xi)$ and $\tilde{\theta}(\xi)= \theta(\xi)+ f(\xi)$, where $f$ is any $\mu$-periodic function, have the same increment.)
Therefore, it is  the differential equation 
  \eqref{A230} that defines $\theta(\xi)$ and not equations \eqref{Y1}-\eqref{Y2}.
   However, once the phase has been determined by solving   \eqref{A230}, we will still need to 
   verify that the corresponding increment reproduces the stationary answer $\Delta(\xi)$ given by  \eqref{Y1}-\eqref{Y2}.

 Using  \eqref{A501} and \eqref{Y1}, we transform     equation \eqref{A230}
  to
  \be
\mu V \theta' =    2 \omega (1- \cos k )  +  \nu  + 2 \sigma I_1 \sin k +   \frac{2I_1 \sin k }{\rho(\xi)}.
\label{A22} 
 \ee 
The integration   of \eqref{A22} gives 
\be 
\label{AA26}
\theta(\xi)  =      \kappa \xi +  \beta {\it \Pi } (\xi),
 \ee
where $\kappa$ and $\beta$ are two coefficients defined by
  \be
\kappa= \frac
{ 2 \omega( 1- \cos k)  + \nu  + 2 \sigma I_1 \sin k}
  {\mu V}
  \label{A260}
  \ee  
  and
  \be
  \beta=  \frac{I_1}{A-B}    \frac{2 \sin k}{\mu V} = \frac{I_1}{A-B}    \times 
  \frac{\omega \ts \dn^3 \mu}{(1+\sigma A)^2 \ts \sn \mu \ts \cn \mu}, 
  \label{A261}
\ee
while $\Pi(\xi)$ is our short-hand notation for Legendre's elliptic integral of the third kind:
\be
{\it \Pi}  (\xi) :=  \Pi \left( \alpha^2; \mathrm{am} (\xi, m); m  \right) = \int_0^\xi \frac{d \eta}{1-\alpha^2 \sn^2 (\eta,m)}.
\label{A603}
\ee
In \eqref{A603}, we have introduced the characteristic
\be 
\alpha^2= \frac{B}{B-A}
\label{ga2} 
\ee
and taken into account that $m< \alpha^2 <1$ when $\sigma=1$ and $\alpha^2 <0$ when $\sigma=-1$. 
We have also introduced a parameter $\nu$ in the definition  \eqref{A260}. 
This is a frequency corrector to be determined.

To fix $\nu$ and finalise 
$\kappa$, we 
subtract the values of the phase 
 \eqref{AA26} with two arguments:
 \be
 \theta_+(\xi)= \theta(\xi+\mu)-  \theta(\xi)= \kappa \mu + \beta \left[ \Pi(\xi+\mu)-\Pi(\xi) \right].
 \label{H1}
 \ee
Using the addition theorems for the elliptic integrals   (see  \ref{Add}) 
 \cite{Cayley,Zhuravskii,Byrd,Olver}, the increment \eqref{H1} is expressed as 
\begin{subequations}
\label{H201} 
\be
        \theta_+(\xi)  
= \kappa \mu + \beta \Pi(\mu)  +\chi(\xi), \label{H200}  \ee
where $\chi(\xi)$ stands for its nonconstant part:
 \be
\chi(\xi) =        s_1  s_{\omega}      \sigma      \arctan q(\xi), \quad 
 q =
 \frac{ \sqrt{\alpha^2(1-\alpha^2)(\alpha^2-m)} \ts    \sn \ts \mu  \ts       \sn(\xi+\mu)  \ts \sn    \ts \xi }
 { 1-\alpha^2 \sn^2 \mu - \alpha^2   \ts \cn \ts \mu   \ts  \dn \ts  \mu   \ts   \sn (\xi+\mu)  \ts \sn \ts \xi}.
 \label{H2} 
 \ee 
 \end{subequations}
 Here we have used the relation
 \[
 \beta  \ts  \sqrt{   \frac{\alpha^2}    {(1-\alpha^2)(\alpha^2-m)}   }  = s_1 s_{\omega},
 \]
 where $s_1= \mathrm{sign}  \, I_1$,
 and $s_{\omega}$ is another sign factor
 ($s_{\omega} = \pm 1$),
 defined by letting 
  \be 
 \omega= s_{\omega} \frac{\cn \mu}{\dn^2 \mu} (1+ \sigma A)^{3/2}.
 \label{omb}
 \ee

 In the defocusing situation $\sigma=-1$,  regardless of the choice of $\alpha^2<0$, 
 and in the focusing case $\sigma=1$,
  provided  $m \leq \alpha^2 < \alpha^2_c$, 
the denominator of the fraction $q(\xi)$ in \eqref{H2}
   is positive for  all values of $\xi$. The proof has been relegated to   \ref{AppendixA}, and 
     $\alpha_c^2$ is also defined there (equations \eqref{Bp9}-\eqref{Bp6}). 
   The expression \eqref{H2}  with those $\sigma$ and $\alpha^2$
   defines a continuous function of $\xi$, for all $\xi \in (-\infty, \infty)$.

By contrast, solutions of the focusing Ablowitz-Ladik equation 
with  $ \alpha^2 > \alpha^2_c$ make the denominator   in \eqref{H2} cross zero as $\xi$ varies.
In that case, the function $\arctan q$ should be considered on a multi-sheet Riemann surface (where the cuts along parts of the imaginary axis outside the segment $(-i, i)$
on each sheet are glued to the opposite cuts
of the neighbouring sheets). The effect of this analytic continuation is that each time the denominator changes from positive to negative, $\pi$ is added to the value of the arctangent in 
\eqref{H2}. Each time the denominator crosses through zero in the opposite direction, $\pi$ is subtracted.

 As we have explained in the beginning of this subsection, 
 the ``travelling" increment  \eqref{H201} must agree to the expression 
 for $\Delta$, the increment afforded by the ``stationary" equations \eqref{Y1} and  \eqref{Y2}. 
  In  \ref{AppendixB} we verify that 
 the difference between 
 $\Delta(\xi)$ and   $\chi(\xi)$ (the variable part of $\theta_+(\xi)$)
  is $\xi$-independent
 and so 
 the two expressions, ``travelling" and ``stationary", 
 can be made equal just by a suitable choice of $\kappa$.

According to equation \eqref{H2},   $\chi(0)$ vanishes.
The identity
\[
\Delta (\xi)  -\chi (\xi)=  \Delta(0)  -\chi (0),
\]
 taken together with  the relation 
  \eqref{Y1}, implies then that the sign of $\sin (\Delta -\chi ) $ agrees with the sign of $I_1$.
That is, when $s_1=1$, the angle difference lies in the range of cotangent: $ 0 \leq \Delta -\chi \leq \pi$, but if $s_1=-1$, it is in the complementary interval:
$ \pi < \Delta -\chi <  2\pi$. 
Equation \eqref{Ap17} in  \ref{AppendixB} gives then 
 \be
\Delta -\chi 
 =  \mathrm{arccot} \left(  \frac{\omega}{I_1} \frac{A + \sigma m \sn^2 \mu}{1+ \sigma A}  \right)  + \frac{\pi}{2}  (1-s_1).
\label{Ap17}
\ee
Finally, imposing the equality $\Delta=  \theta_+$, where $\theta_+$ is the ``travelling" increment  \eqref{H201}, 
  we recover $\kappa \mu$:
 \bea
\nonumber 
 \kappa \mu=  & \mathrm{arccot}    \left(  \frac{\omega}{I_1} \frac{A + \sigma m \sn^2 \mu}{1+ \sigma A}\right)   - \beta \Pi (\mu)+\frac{\pi}{2} (1-s_1) 
 \\
 = &  \mathrm{arctan}    \left(  \frac               {I_1}       {\omega}
 \frac{1+ \sigma A}                          {A + \sigma m \sn^2 \mu}      \right) 
   - \beta \Pi (\mu)+    \frac{\pi}{2} (1-s_{\omega}).
   \label{S2} 
    \eea

\subsection{Swinging wave solution}

     \begin{figure*}[t!]
    \centering
    \begin{subfigure}{0.495\textwidth}
              \includegraphics[width=\textwidth]{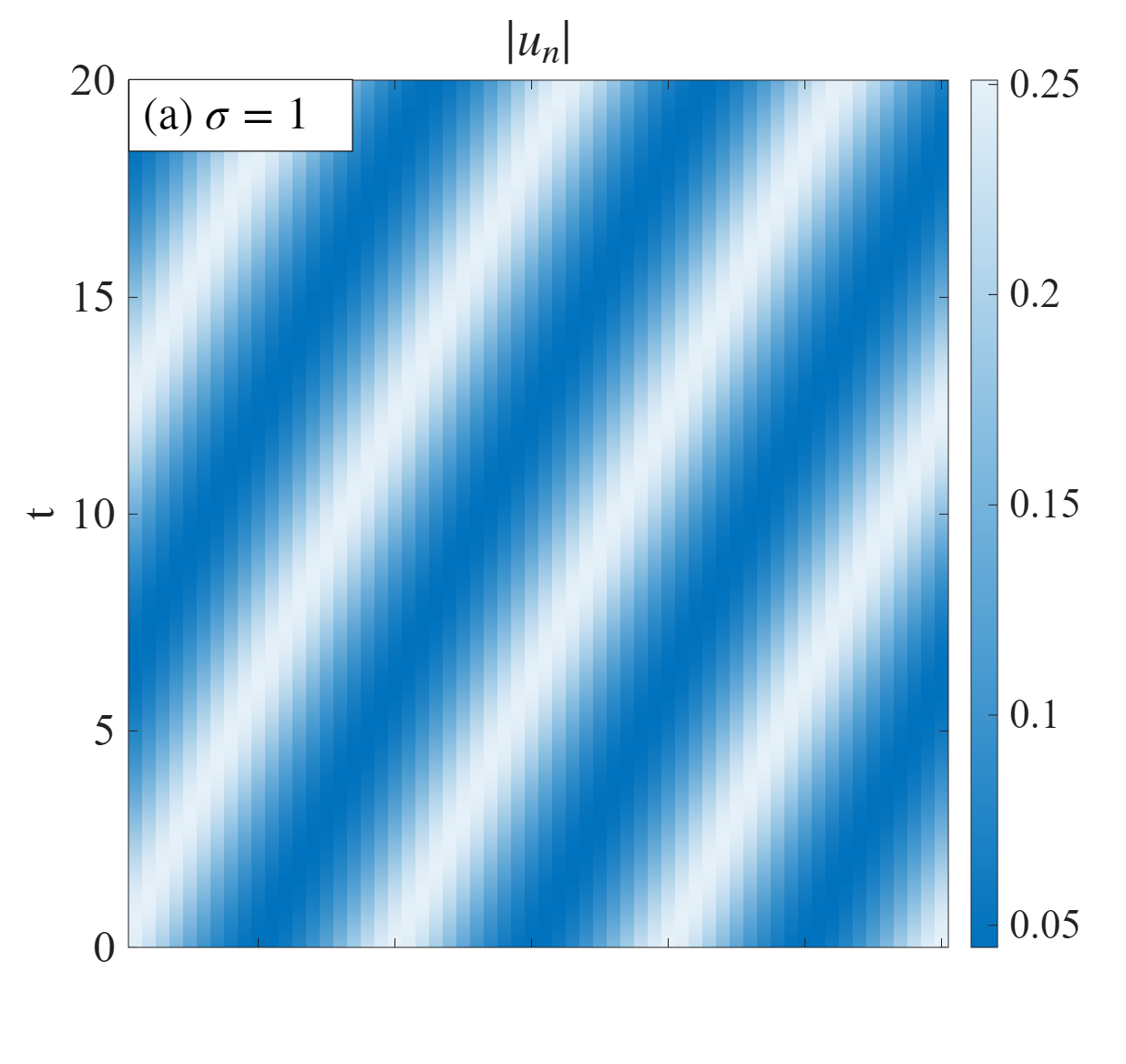}
  
    \end{subfigure}\hfill
    \begin{subfigure}{0.495\textwidth}
                \includegraphics[width=\textwidth]{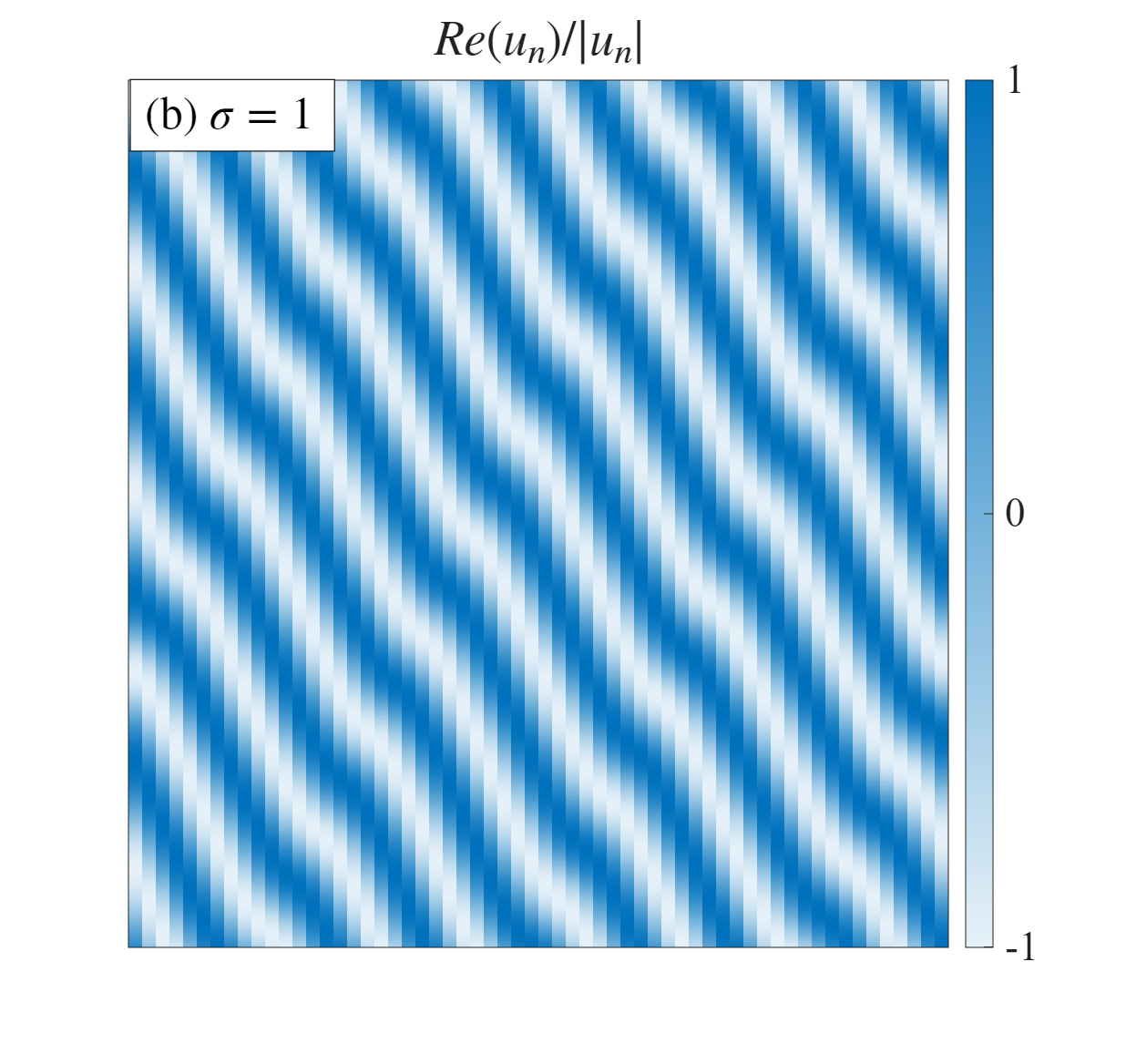}
    \end{subfigure}\hfill
    \centering
    \begin{subfigure}{0.495\textwidth}
                \includegraphics[width=\textwidth]{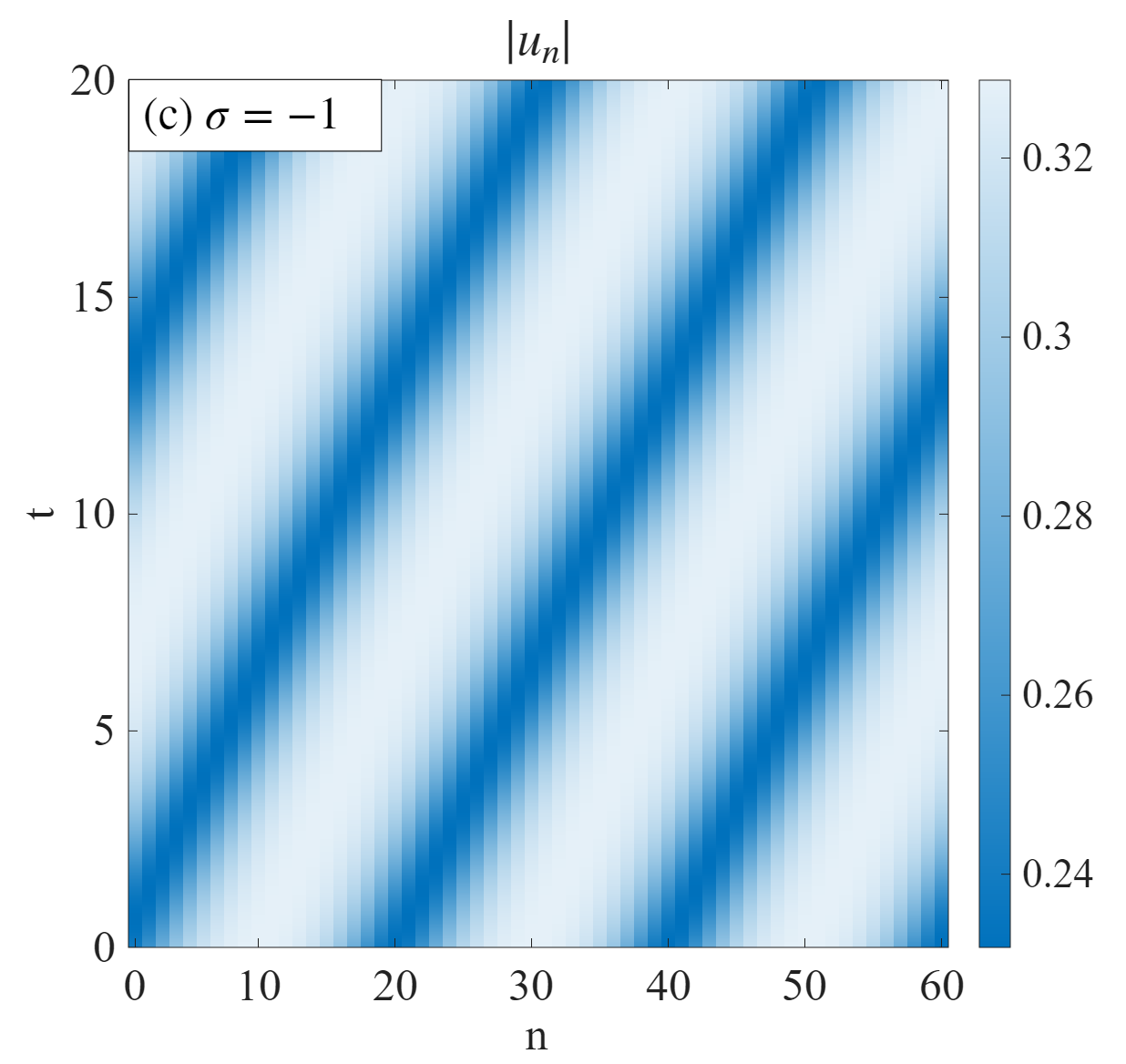}
    \end{subfigure}
    \begin{subfigure}{0.495\textwidth}
                \includegraphics[width=\textwidth]{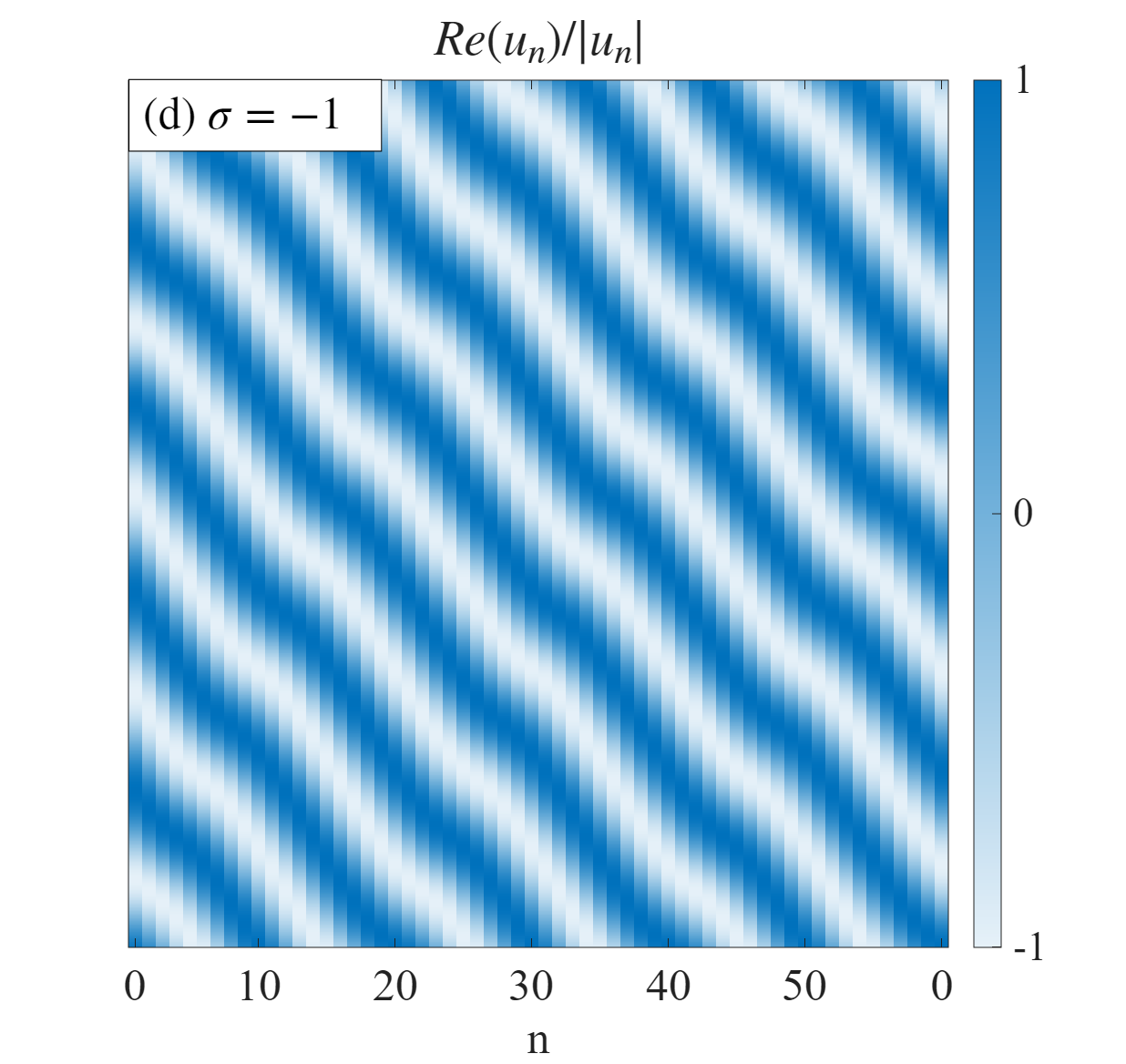}
    \end{subfigure}
    \caption{Nonlinear-phase wavetrains of the focusing (a,b) and defocusing (c,d) Ablowitz-Ladik equation \eqref{A1}. In both cases we have chosen 
    $I_1>0$ and $\omega>0$, 
    with $N=60,\;m=0.9,\;\mu=6K(m)/N$, and $k=3\pi/10$.   
    The same amplitude values  as in Fig \ref{fig:stat_foc_defoc} are used:
    $A=2.003\times 10^{-3}$    
     for $\sigma=1$ and $A=1.080\times 10^{-1}$
     for $\sigma=-1$. 
    (Note that the  resulting winding numbers   are   {\it not\/}  the same as the winding numbers in Fig   \ref{fig:stat_foc_defoc}.
    The wavetrains in (a,b) satisfy $\Theta_N-\Theta_0=20 \pi$;  those in (c,d) have $\Theta_N-\Theta_0=24 \pi$.)
            The time interval between the phase rolls  (panels (b) and (d)) exhibits periodic shortening; 
    the wave ``swings".   
        }
    \label{fig:travelling_wave}
\end{figure*}

Combining the above results, the travelling wavetrain solution of  the Ablowitz-Ladik equation \eqref{A1} assumes the form
\begin{subequations} \label{final_sol}
\begin{align}
    U_n(t)=\sqrt{A-B \ts \cn^2(\xi_n,m)}    \ts  
    e^ {i \Theta_n},   \label{K1}    \\
    \Theta_n= 
      \beta\Pi(\xi_n)  +
     (\kappa \mu +k)n
    + 2   (\omega\cos k-\sigma I_1\sin k)t.
    \label{K2}
\end{align}    
\end{subequations} 
Here $\xi_n$ is the travelling-wave combination defined by \eqref{xin} and the elliptic integral $\Pi(\xi)$ is given by \eqref{A603},
with the characteristic $\alpha^2$ as in \eqref{ga2}.
Letting $\omega = s_{\omega} \tilde{\omega}$ and $k= \tilde{k} -\frac{\pi}{2}(1-s_{\omega})$ eliminates the sign factor $s_{\omega}$ from the solution, 
together with the  $\frac{\pi}{2}(1-s_{\omega})$  term
in \eqref{S2}. 
This means that without loss of generality, the parameters $\omega$ and $\kappa \mu$ may be assigned the following values in equation \eqref{final_sol}:
\be 
 \omega=  \frac{\cn \mu}{\dn^2 \mu} (1+ \sigma A)^{3/2},
 \quad 
  \kappa \mu= 
   \mathrm{arctan}    \left(  \frac               {I_1}       {\omega}
 \frac{1+ \sigma A}                          {A + \sigma m \sn^2 \mu}      \right) 
   - \beta \Pi (\mu).
   \label{Y40}
   \ee

The discretisation stepsize $\mu$,  elliptic modulus $m$,  amplitude   $A$,
wavenumber $k$, and centering parameter $x_0$ can be chosen arbitrily,
within the appropriate limits. The velocity $V$ and the coefficient $B$ are
 as in \eqref{A26} and \eqref{F5}, respectively; the quantity   $I_1$ is given by $s_1 |I_1|$, with $I_1^2$ as  in \eqref{F7}, while the coefficient
\be 
\beta= s_1 \sqrt{
\frac{(1-\alpha^2)(\alpha^2-m)}{\alpha^2}.
}
\ee

Figure \ref{fig:travelling_wave} exemplifies travelling waves with $\sigma=1$ and $\sigma=-1$. 
The modulus-squared of the solution of  the focusing equation (Fig \ref{fig:travelling_wave} (a))
 represents a comb of pulses on a finite-height shelf, while
    in the defocusing case (Fig \ref{fig:travelling_wave} (c)) the wave-train consists of a sequence of  ``dark spots" moving against a uniformly ``illuminated"  background.
    
 The panels (b) and (d) illustrate the propagation of the phase of the solution. The phase velocity (the slope of the rolls) experiences  periodic  increases, or swings, 
 resulting from the nonlinear term $\beta \Pi(\xi_n)$ in \eqref{K2}.

 Letting $k=0$ in equation \eqref{A26} produces $V=0$.
 Thus, the expression \eqref{final_sol} with $k=0$  provides a stationary wavetrain solution of the Ablowitz-Ladik equation  --- a 
  closed-form alternative to the solution by quadrature  obtained in section \ref{Varied_Phase}.  The phase  of the complex quantity $u_n= \sqrt{\rho_n}e^{i \theta_n}$
  is given by
  \be 
  \theta_n =\beta  \left[ \Pi(\mu n + {x}_0 ) - n \Pi(\mu) 
  \right] +  n  \ts  \mathrm{arctan}    \left(  \frac               {I_1}       {\omega}
 \frac{1+ \sigma A}                          {A + \sigma m \sn^2 \mu}      \right), 
   \ee
   and its absolute value squared is
   \[
   \rho_n= A-B \cn^2(\mu n + {x}_0, m).
   \]

\subsection{Continuum limit}

Letting $\psi_n(\tau)= \psi(x_n, \tau)$,    where $x_n=\mu n +x^{(0)}$;   
sending $\mu \to 0$, and assuming that the function $\psi(x, \tau)$ and its partial derivatives  $\partial^j \psi/ \partial x^j$ are of order $\mu^0$,
the Ablowitz-Ladik lattice \eqref{B1} reduces to a partial differential equation:
\be
i \psi_\tau + \psi_{xx} + 2 \sigma |\psi|^2 \psi=0.
\label{C1}
\ee
The cnoidal wave solution of the nonlinear Schr\"odinger equation \eqref{C1} takes the form
$ \psi(x, \tau)= \sqrt{r} e^{i (\varphi+  \tilde \omega \tau)}$,  where $r$ and $\varphi$ are given by equations \eqref{NLS12} and \eqref{NLS16} 
in \ref{cnoNLS}.

One can readily obtain the continuum limit of the travelling wavetrain  \eqref{final_sol}. 
Letting $A= \mu^2 a$,   $V=\mu \varv$, and sending $\mu \to 0$, we verify that the 
expression $\mu^{-2}|U_n|^2$, with $U_n$ as in \eqref{K1}, reduces to the modulus squared of the nonlinear Schr\"odinger solution,  equation \eqref{NLS12}.

Turning to the phase of the lattice solution, equation \eqref{K2}, we observe that as $\mu \to 0$, its parameters satisfy
\be
\nonumber
\beta \to    s_1 \sqrt{ \frac{a(m'-\sigma a)}{a+\sigma m}},
\quad 
 \alpha^2 \to {\tilde \alpha}^2,
\ee
where ${\tilde \alpha}^2$ is defined in \eqref{NLS18}.  Without loss of generality,  we may identify $s_1= \tilde s_1$;
then the $\Pi$ term in \eqref{K2} reduces to its counterpart in \eqref{NLS16}. Furthermore, the invariant $I_1$ reduces to 
$ \mu^3 J_1$, with $J_1$ as in \eqref{NLS21}; in view of this relation,
 equation \eqref{S2} yields  $\kappa \mu = O(\mu^2)$, implying that the $\kappa \mu$ term does not contribute to the lattice phase expression in the limit.
Having let  $k=(\varv/2)\mu$ and substituted $\xi_n+ \varv \tau- x^{(0)}$ for $\mu n$, we note that 
the $\xi_n$-component of the expression $\Theta_n-2t$  transforms into the $\eta$-component of the continuum phase $\varphi+\tilde \omega \tau$.
Finally,  expanding $\omega$ and $\cos k$  in the Taylor series up to $O(\mu^2)$
and
using    
 \eqref{NLS13}, we verify that the time component of the lattice phase reduces to the temporal component of 
its continuum analogue as well.

In summary, the continuum limit of the swinging wave \eqref{final_sol} is given by 
the cnoidal wave of the nonlinear Schr\"odinger equation:
equations \eqref{NLS12} and \eqref{NLS16}.

\section{Solitons, bright and dark}
\label{solitons}

Of special interest  is the  limit of infinite period: $m=1$. 

In the focusing case ($\sigma=1$), 
sending $m \to 1$ brings the background amplitude, $A$, to zero. Equations \eqref{final_sol} reduce
to the well-known
bright soliton (with a linearly changing phase):
\begin{equation}
    U_n(t)=\sinh \mu  \ts \frac{  e^{i    \left[kn+2\cosh(\mu)\cos (k) t\right]  }    }{   \cosh \left[ \mu(n-Vt) +  x_0 \right]}, 
\end{equation}
where
\begin{equation}
    V= 2 \frac{\sinh \mu }{\mu} \sin k.
\end{equation}

Solutions of  the defocusing  ($\sigma=-1$)
Ablowitz-Ladik equation prove to be less familiar.
As $m \to 1$, 
 the elliptic integral reduces to a sum of elementary functions:
\be
\Pi(\xi) =
\Pi \left( \alpha^2; \mathrm{am} (\xi, 1); 1  \right)
=\frac{\xi + \delta \arctan (\delta \tanh \xi)}{1+\delta^2},
\label{Q1}
\ee
where $\delta = \sqrt{-\alpha^2}$. (We remind that $\alpha^2$ is negative  in the defocusing case.)
The value of $\delta$ pertinent to  our solution \eqref{final_sol}
is given by  \eqref{ga2}:
\be
 \delta = \frac{ \sqrt{1-A} \tanh \mu}{\sqrt{A- \tanh^2 \mu}}.
\label{Y15} \ee
Other parameters in \eqref{final_sol} are also expressible in elementary functions. 
Letting $m=1$ in equation    \eqref{A261} we obtain
\be
\frac{\beta}{1+ \delta^2} = s_1 
 \frac          {\sqrt{A- \tanh^2 \mu}}  { \sqrt{1-A} \tanh \mu},
 \label{Y10}
\ee
while the argument of the arctangent in \eqref{Y40} becomes
\be 
    \frac               {I_1}       {\omega}
 \frac{1+ \sigma A}                          {A + \sigma m \sn^2 \mu}    =  \frac{s_1 }{A}  \sqrt{(1-A)(A- \tanh^2 \mu)}.
 \label{Y11}
 \ee
Making use of \eqref{Q1}, \eqref{Y10} and \eqref{Y11},  the expression \eqref{Y40} simplifies:
\be
\kappa \mu = s_1   \arctan \sqrt{ \frac{A- \tanh^2 \mu}{1-A}} 
-s_1  \sqrt{ \frac{A- \tanh^2 \mu}{1-A}} \frac{\mu}{\tanh \mu}          .
\label{G11}
\ee
We also note that 
\be
\sqrt{ \rho(\xi_n)} e^{i s_1 \arctan (\delta \tanh \xi_n)}= \sqrt{A-B}  (1+ i s_1  \ts  \delta \tanh \xi_n).
\label{Y12}
\ee

Finally, substituting the sum \eqref{G11} in  our cnoidal-wave solution  \eqref{final_sol} 
and making use of \eqref{Q1} and \eqref{Y12}, 
we arrive at the expression for the travelling dark soliton:
\be 
U_n(t)=\cosh \mu \sqrt{ A-\tanh^2 \mu}  \ts (1+ i s_1 \ts  \delta \tanh \xi_n)  e^{i (Qn+ \Omega t)}.
\label{Y14}
\ee
Here, the carrier's wavenumber is given by
\be
Q=k + s_1 \arctan  \sqrt{ \frac{A-\tanh^2 \mu}{1-A}},
\label{Y50} 
\ee
 and the corresponding frequency is 
 \be
\Omega= 2 (1-A) \cos Q.
\ee
We remind that  $\xi_n$ is the travelling wave combination, $\xi_n= \mu(n-Vt)+ x^{(0)}$, 
where $V$ is given by equation \eqref{A26} with $\omega$ as in \eqref{omb}:
\be 
V= 2  \sqrt{1-A}  \ts \frac{\sinh \mu}{\mu} \sin k.
\label{Y16}
\ee
The coefficient  $\delta$ is as in \eqref{Y15}, while 
 the parameters $\mu$, $A$
$(\tanh^2 \mu \leq A \leq 1)$, $k$ and ${x}_0$,  as well as the sign factor $s_1 =\pm 1$, may be chosen arbitrarily.

The dark soliton represents a region of low intensity $|U_n|^2$ propagating 
at velocity $V$ against the background of uniform illumination,
where the background wave is itself non-stationary. 
In the continuum nonlinear Schr\"odinger equation, the
 moving background (with a soliton travelling at its own speed relative to the background)
 can be obtained just by the Galilean transformation. By contrast, in
a discrete system, localised structures propagating over the moving wave constitute an independent family with no obvious relation to dark solitons 
travelling over a stationary background.

The special situation of stationary background corresponds to 
 $Q=0$ in \eqref{Y14}. 
The velocity of the soliton in this case  is straightforward from \eqref{Y50} and \eqref{Y16}:
\be 
V= - s_1  \frac{\sinh 2 \mu}{\mu} \sqrt{(1-A)(A- \tanh^2 \mu)}.
\ee
As in the defocusing nonlinear Schr\"odinger equation, the amplitude and velocity of the Ablowitz-Ladik dark soliton are related.


\section{Periodic and Quasiperiodic Solutions}\label{sec:numerical_implementation}
\label{periodic} 
\subsection{Travelling periodic waves}
\label{trav_per}

A physically meaningful subclass of solutions comprises waves on a ring: a finite chain with joined ends.
 These have to satisfy the periodicity conditions 
$ U_{n+N}=U_n$, where $N$ is the number of sites in the chain.
The periodicity of the complex variable $U_n= \phi_n e^{i \Theta_n}$  requires $\phi_{n+N}=\phi_n$
and constrains its phase.

 The amplitude $\phi_n$ is periodic provided a natural number $\ell $ of periods of the cnoidal wave matches the length of the chain:
$2 K(m)   \times \ell= \mu N$.   
This equation  defines a set of admissible values of the chain spacing in the interval 
  $0 < \mu <2K(m)$:
  \be 
  \mu= \mu_\ell := \frac{ 2 \ell}{N} K(m),   \quad \ell = 1, 2, ..., N.
  \label{Y21}
  \ee
  (The index $\ell$  can be visualised    as the number of wave crests on the ring.)

The constraint the phase should satisfy is
\be 
\label{Y30}
\Theta_{n+N} = \Theta_n + 2 \pi w,
\ee
where $w$ (``winding number") is an integer. 
Substituting $\Theta_n$ from \eqref{K2} gives
\be 
\beta   \left[  \Pi(\xi_{n+N}) -   \Pi(\xi_n)  \right]= 2\pi w -(\kappa \mu+k)N.
\label{Y31}
\ee
Both terms in the left-hand side of \eqref{Y31} are time-dependent; however the time derivative of the bracketed expression  vanishes for $\mu=\mu_\ell$. 
Therefore the left-hand side maintains its value for all $t$ --- in particular, for $t$ rendering $\xi_n=0$:
\be
\Pi(\xi_{n+N})- \Pi(\xi_n)=  \Pi(\mu_\ell N).
\label{Y32}
\ee
Since $\mu_\ell N =  2 \ell K(m)$, the right-hand side  in \eqref{Y32} equals  $\Pi(2 \ell K(m))= 2 \ell  \ts \Pi(\alpha^2, m)$, where $\Pi(\alpha^2, m)$ is the complete 
elliptic integral of the third kind:
\[
\Pi(\alpha^2, m) = \int_0^{K(m)}  \frac{d \eta}{1-\alpha^2 \sn^2 (\eta,m)}.
\]
Solving \eqref{Y31} for $k$ gives then
\be
k= \beta \left[ \Pi(\mu_\ell) -\frac{2 \ell}{N} \Pi(\alpha^2, m) \right] - \arctan \left( \frac{I_1}{\omega} \frac{1+ \sigma A}{A+ \sigma m \sn^2 \mu_\ell}  \right) + 2 \pi \frac{w}{N},
\label{Y33}
\ee
$w \in \mathbb{Z}$. 
To avoid any notational misinterpretation, we remind that  $\Pi(\mu_\ell)$ in the line above stands for the {\it incomplete\/} elliptic integral:
\[
 \Pi(\mu_\ell) = \Pi\left(\alpha^2; \mathrm{am}(\mu_\ell,m); m \right)=  \int_0^{\mu_\ell} \frac{d \eta}{1-\alpha^2 \sn^2 (\eta,m)}.
 \]

For each  $A$, $m$,  $N$ and $\ell$, equation \eqref{Y33} defines a set of wavenumbers $k=k_{w} (A,m, N, \ell)$ --- and, by virtue of equation  \eqref{A26}, --- a set of 
velocities $V_{w}(A,m, N, \ell)$, where $w=1, 2, ..., N$.  
This means that in the Ablowitz-Ladik  necklace of $N$ elements, 
the wave pattern with an amplitude $A$, $\ell$ crests and the spacing  between the crests given by $2K(m)$
can travel at any  of $N$ nonequivalent velocities.

\subsection{Standing wave quantisation} 
\label{stanqua}

Of particular interest are  stationary  ($V=0$) periodic patterns.

\begin{figure*}[t!]
    \centering  
        \begin{subfigure}{0.495\textwidth}
                \includegraphics[width=\textwidth]{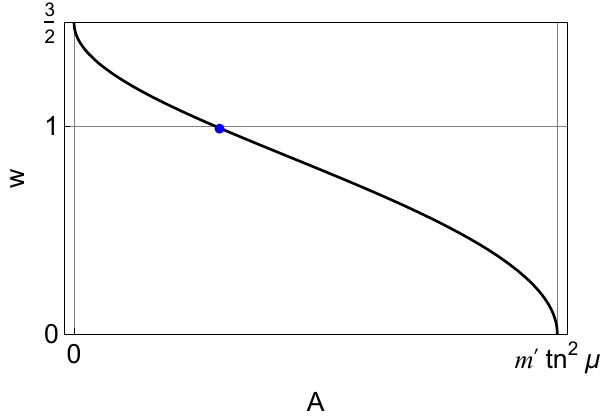}
    \end{subfigure}\hfill
    \begin{subfigure}{0.495\textwidth}
                \includegraphics[width=\textwidth]{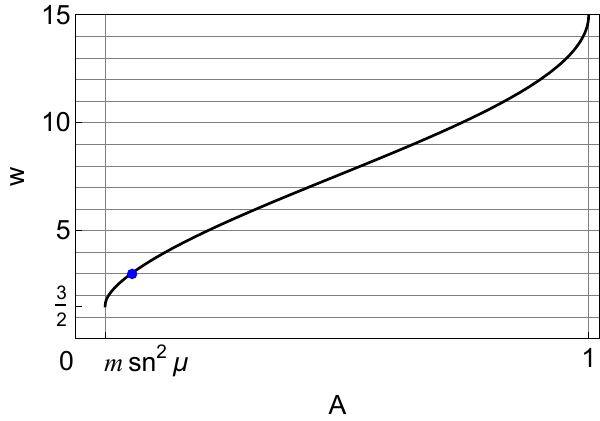}
    \end{subfigure}
    \hfill     
    \caption{The right-hand side of equation   \eqref{Y53} for a fixed $m$ and varied $A$.
         The left panel pertains to $\sigma=1$ and the right one to $\sigma=-1$; in either case $N=60$, $m=0.9$ and  $\ell=3$. 
         Intersections of the $w(A)$ curve with the  horizontal gridlines (corresponding to the  integer values of $w$)
    give 
    the quantised standing wave amplitudes, $A_w$.
    Thus, the horizontal coordinate of the blue blob in (a) is $A_1$ and the coordinate of a similar blob in (b) is $A_3$.
           The notation $m^\prime \mathrm{tn}^2 \mu$ in the left panel  stands for $(1-m) \sn^2 (\mu_\ell, m) /  \cn^2 (\mu_\ell, m)$. In the right panel, 
        $m\ts \sn^2 \mu= m \sn^2(\mu_\ell, m)$.
         }
    \label{Aw}
\end{figure*}

Waves travelling with zero velocity are selected by choosing $k=0$ in equation \eqref{A26}. On the other hand, 
letting $k=0$ in
equation \eqref{Y33} allows one to express $w$ as a function of $A$ and $m$:
\be
 w
= \frac{ \beta N}{2\pi}   \left[ \frac{2 \ell}{N} \Pi(\alpha^2, m) -
\Pi(\mu_\ell)  \right] +\frac{N}{2 \pi}  \arctan \left( \frac{I_1}{\omega} \frac{1+ \sigma A}{A+ \sigma m \sn^2 \mu_\ell}  \right).
\label{Y53}
\ee
For a generic pair of $m$ and $A$ the expression  \eqref{Y53} does not  assume an integer value --- and so there are no stationary periodic solutions with this pair of parameters. 
However for a given elliptic modulus $m$   there is a particular set of integers $w$ and
amplitudes  
 $A_w(m)$
 such that the right-hand side of \eqref{Y53} with $A=A_w$
equals  $w$. The quantised amplitudes $A_w$ characterise  a set of  standing waves on an $N$-site ring of the length $\mu N$, with $\ell$ crests.
Figure \ref{Aw} illustrates the graphical determination of  $A_w$.

\subsection{Quasiperiodic solutions}
\label{quasi}

The approach of  section \ref{trav_per} can be trivially extended  to the quasiperiodic boundary condition
\begin{equation}
    U_{n+N}=U_{n}e^{i\gamma},    \label{Y34} 
\end{equation}
with $0<\gamma<2\pi$.

Indeed, choosing $\mu$ to be given by $\mu_\ell$,  one of the values in  \eqref{Y21}, 
we make sure that the absolute value of $U_n$ is $N$-periodic.
The boundary condition 
\eqref{Y34}   translates   then  into the following constraint on 
 the phase of the solution  \eqref{final_sol}:
\be
\Theta_{n+N}= \Theta_n + \gamma + 2 \pi \tilde{w},
\label{Y35}
\ee
where $\tilde w$ is an integer. This is equivalent to  \eqref{Y30} with $w= \gamma/ (2 \pi) + \tilde{w}$.
All we need to do to satisfy \eqref{Y35} is 
let $k$ be defined by 
 equation \eqref{Y33} with $w= \gamma/ (2 \pi) + \tilde{w}$.

\section{Concluding Remarks}
\label{Conclusions}

We have constructed novel classes of travelling-wave solutions of the focusing and defocusing Ablowitz-Ladik equation.
The modulus  of each complex solution is expressible in terms of the Jacobi $\cn$ function while its phase (complex angle) involves the Legendre's elliptic integral of the third kind.
Unlike cnoidal waves obtained by earlier authors,  the  phase variable has a nonlinear dependence on time.
The wave ``swings"; see Fig \ref{fig:travelling_wave} (b,d).

As its wavelength (the period of the Jacobi function) tends to infinity, the travelling wave transforms to a soliton, bright or dark. 
The emerging bright soliton  of the focusing equation is extensively documented 
in the literature while the dark soliton solution of its defocusing counterpart is new. The latter object represents a localised depression 
 travelling over a non-stationary exponential wave background.
 
 Lattice solitons in generic models may propagate without exciting radiation and thus defying the Peierls-Nabarro barrier  \cite{ZES,Flach,CK,PR1,IP}.
 The corresponding special, isolated, propagation speeds depend on the discretisation parameter and 
 are referred to as the sliding velocities  \cite{OPB,BvH,AMM}. 
 The  Ablowitz-Ladik solitons determined in
 section \ref{solitons} can propagate with an arbitrary speed; thus any $V$ is a sliding velocity here.
This property  is not necessarily related with the integrability of the Ablowitz-Ladik equation; 
there are nonintegrable lattice models whose solitons can slide at an arbitrary speed \cite{BvH}.

Despite its  modulus being expressible in terms of a periodic function, the wave travelling at a generic speed is  neither periodic nor quasiperiodic.
The wave can  be seen as circulating around a ring of $N$ sites only if it moves with one of special velocities.
These velocities form a  set  of $N$ values that is determined by the amplitude of the wave, its length and the inter-site spacing.

We conclude with three remarks  on the possible relevance of our new Ablowiz-Ladik solutions to other lattice systems. 

First, the transformation $q_n(t)= e^{-i  n\varphi } U_n(t)$, where $\varphi$ is a constant angle, 
takes  solutions of  the Ablowitz-Ladik equation \eqref{A1}
 to those  of the  discrete Hirota equation \cite{narita_1990} 
\begin{equation}
\label{Hirota}
    i  \frac{ \partial q_n }{\partial t}  + \left(           e^{i\varphi}           q_{n+1}+                    q_{n-1}                  e^{-i  \varphi}         \right)(1+\sigma |q_n|^2)=0.
\end{equation}
In particular, the quasiperiodic  Ablowitz-Ladik 
travelling waves  constructed in section \ref{quasi}  and 
satisfying the boundary condition \eqref{Y34}  with $\gamma= N \varphi$ 
are mapped to periodic solutions of \eqref{Hirota} with $q_{n+N}= q_n$.

Choosing $\varphi = \pi/2$, we obtain  periodic waves of 
the complex discrete modified KdV equation \cite{ChenPeli_1}
\begin{equation}
   \frac{\partial q_n}{\partial t} + \left(q_{n+1} - q_{n-1})\right(1+ \sigma |q_n|^2)=0.
\end{equation}

Second, 
the swinging-wave solutions may  be deformed by homotopy continuation from the Ablowitz-Ladik lattice to its nonintegrable counterpart, 
the discrete nonlinear Schr\"odinger equation:
\be
i \frac{\partial U_n}{\partial t} + U_{n-1} + U_{n+1}   +  2 \sigma |U_n|^2U_n=0.
\label{dnls}
\ee
The latter system plays an immense role in physical applications, most notably in optical and matter-wave settings \cite{HT,EJ,kevrekidis_discrete_2009}.
Previous studies  analysed its solutions 
as perturbations of the Ablowitz-Ladik 
linear-phase  soliton  and breather
\cite{cai1,cai2, melvin_discrete_2009, sullivan_kuznetsov-ma_2020, Mithun1,Hennig1,Hennig2,Mithun2}. 
The homotopy continuation of the new cnoidal-wave and soliton solutions 
of the Ablowitz-Ladik equation should yield  nontrivial-phase
solutions of equation     \eqref{dnls}.

Finally, it is worth noting that the Ablowitz-Ladik equations describe the integrable motions of a discrete curve 
on the sphere \cite{Doliwa} and the asymptotic behavior of the temperature correlation function of the quantum XY spin chain \cite{Its}. 
New classes of solutions may admit interpretation in these 
and similar contexts.


\paragraph{Acknowledgments}
We thank Nora Alexeeva for   assistance  with computations.
This research was supported by the NRF of South Africa (grant No  SRUG2204285129).

\appendix

\section{Addition theorem for elliptic integral} 
\label{Add} 

Cayley \cite{Cayley} casts the addition theorem into the following simple form:
\be
 \Pi(\xi_1+\xi_2)  - \Pi (\xi_1) - \Pi(\xi_2) 
= \int_0^R \frac{dr}{1+ \mathcal A r^2}.
\label{Ad1} 
\ee
Here $\Pi(\xi)$ is our short-hand notation for Legendre's elliptic integral of the third kind, equation \eqref{A603}. The upper limit of integration in \eqref{Ad1} 
is given by
\be
R=  \frac{\alpha^2   \ts \sn  \ts \xi_1 \ts  \sn  \ts   \xi_2 \ts    \sn (\xi_1+ \xi_2)}
{1-\alpha^2 + \alpha^2   \ts  \cn   \ts  \xi_1    \ts    \cn   \ts  \xi_2   \ts    \cn (\xi_1+\xi_2)},
\nonumber 
\ee
and the parameter $\mathcal A$ is defined by
\be
\mathcal A   := (1-\alpha^2) \left( 1- \frac{m}{\alpha^2} \right).
\nonumber
\ee
Note that $\mathcal A>0$ regardless of whether $\alpha^2<0$ or
$m < \alpha^2 <1$.

Doing the integral in \eqref{Ad1} we arrive at two cases of the addition theorem, different in the sign in front of the right-hand side:
\bea
 \Pi(\xi_1+\xi_2)  - \Pi (\xi_1) - \Pi(\xi_2) =
 \pm \sqrt{\frac{\alpha^2}{(1-\alpha^2)(\alpha^2-m)}}   \nonumber \\
\times   \arctan \left[    \frac{ 
 \sqrt{ \alpha^2(\alpha^2-m)(1-\alpha^2)}  \ts
\sn  \ts \xi_1 \ts  \sn  \ts   \xi_2 \ts    \sn (\xi_1+ \xi_2)}
{1-\alpha^2 + \alpha^2   \ts  \cn   \ts  \xi_1    \ts    \cn   \ts  \xi_2   \ts    \cn (\xi_1+\xi_2)}  \right].
\label{Ad2} 
\eea
The top (positive) sign in \eqref{Ad2} corresponds to $\alpha^2>0$ and the bottom (negative) one to $\alpha^2<0$. 

The occurrence of two signs has a simple explanation.  When $\alpha^2>0$, the integral $\Pi(\xi)$ grows faster than $\xi$,
whereas when $\alpha^2<0$, it grows slower than linearly. Therefore, $\Pi(\xi_1+\xi_2)$ is greater than $\Pi(\xi_1)+\Pi(\xi_2)$ in
the former case and less in the latter. 

There is some confusion in the reference literature about the correct form of the addition theorem.
Zhuravskii \cite{Zhuravskii} writes the rule \eqref{Ad2} with only a negative sign on the right-hand side,  but does not make clear whether he considers  the case $\alpha^2<0$ exclusively. 
Byrd and Friedman \cite{Byrd} also give \eqref{Ad2} with only a negative sign, while erroneously claiming that their addition result is valid for either sign of $\alpha^2$.
Finally, the most recent resource \cite{Olver} presents the addition theorem in a form that is exactly reducible to equation \eqref{Ad2}.

\section{Travelling increment, regular and singular}
\label{AppendixA}

As the coordinate $\xi$ is varied,  the 
argument of the arctangent in \eqref{H2} may develop singularities.
In such a case, the
 equation \eqref{H2} will not
define a continuous function of $\xi$. 
In  this Appendix we characterise the parameters $\sigma$ and $\alpha^2$ that give rise to the singular behaviour.

Let $\mathcal D$ denote the denominator of the fraction $q(\xi)$       in  \eqref{H2}:
\be
\mathcal D (y)= 1 - \alpha^2   \left[  \sn^2 \mu +  y \ts   \cn \mu \ts \dn \mu  \right],
\label{Bp1} 
\ee
where
\[
 y=y(\xi)= \sn \ts  \xi   \ts  \sn (\xi+\mu).
\]
As $ \xi$ varies from $-\infty$ to $\infty$, 
the $2K(m)$-periodic function $y(\xi)$ oscillates between  $y_{\mathrm{min}} =- \sn^2 \lambda$ 
and  $y_{\mathrm{max}}= \cn^2 \lambda/  \dn^2 \lambda$, and 
the linear function $\mathcal D(y)$ 
maps the interval $(y_\mathrm{min}, y_\mathrm{max})$ onto  $(\mathcal D_{\mathrm{min}}, \mathcal D_{\mathrm{max}})$.
To determine whether the argument of the arctangent in  \eqref{H2} is singular or not, we need to determine whether the 
range $(\mathcal D_{\mathrm{min}}, \mathcal D_{\mathrm{max}})$ contains the origin.

In the  case of the defocusing Ablowitz-Ladik equation ($\sigma=-1$) the parameter  $\alpha^2$ is negative.   
When $\cn \mu>0$, the lowest value of $\mathcal D(y)$  is attained at $y=y_{\mathrm{min}}$:
\be
\mathcal D_\mathrm{min}  = 1 - \alpha^2   \left( 
 \sn^2 \mu -  \cn \mu \ts \dn \mu   \ts   \sn^2 \lambda \right). 
\label{Bp5} \ee
Using identities satisfied by the Jacobi functions, this expression is cast in the form with a positive lower bound:
\[
\mathcal D_\mathrm{min}  =1- \alpha^2 \frac{(1- \cn \mu)(1+ \cn \mu+ \dn \mu)}{1+ \dn \mu} \geq 1.
\]
On the other hand, when $\cn \mu<0$, the function $\mathcal D(y)$ reaches its minimum at  $y=y_\mathrm{max}$:
\be
\mathcal D_\mathrm{min} = 1 - \alpha^2 \left(  \sn^2 \mu + \cn \mu  \ts \dn \mu    \frac{\cn^2 \lambda}{\dn^2 \lambda}  \right).
\label{Bp2} 
\ee
Transforming to functions of double argument, this becomes
\be
\mathcal D_\mathrm{min} = 1 - \alpha^2 \left[   (1-m) \sn^2 \mu + \cn \mu + \dn \mu \right].
\label{Bp3} \ee
The bracketed expression  in \eqref{Bp3} is positive; hence $\mathcal D_\mathrm{min} \geq 1$.

Turning to the focusing case  ($\sigma=1$), the parameter $\alpha^2$ becomes positive  ($m \leq \alpha^2 \leq 1$). 
When $\cn \mu>0$, the maximum  of $\mathcal D(y)$ for the fixed $\alpha^2$ is given by
\[
\mathcal D_\mathrm{max} = 1- \alpha^2 \left(
\sn^2 \mu - \cn \mu \ts \dn \mu  \ts  \sn^2 \lambda \right).
\]
Transforming to the double argument, this becomes
\[
\mathcal D_\mathrm{max} = 1- \alpha^2 \frac{1- \cn \mu}{1+ \dn \mu} (1+ \dn \mu + \cn \mu).
\]
The lowest value of  $\mathcal D_\mathrm{max} $ over all admissible $\alpha^2$ corresponds to  $\alpha^2=1$. 
One readily checks that this value is positive:
\[
\left.  \phantom{\frac12}   \mathcal D_\mathrm{max}  \right|_{\alpha^2=1}= \frac{\cn \mu(\dn \mu+ \cn \mu)}{1+ \dn \mu} >0.
\]
Therefore, $\mathcal D_\mathrm{max}$ is positive regardless of the choice of $\alpha^2$. 
On the other hand, the minimum  of $\mathcal D(y)$ for the fixed $\alpha^2$ is given by \eqref{Bp2}
which we write as 
\be
\mathcal D_\mathrm{min} = 1 - \alpha^2 \left[ 1 + \frac{\cn \mu (\dn \mu- \cn \mu)}{1+ \dn \mu}  \right].
\label{Bp7} 
\ee

The minimum value \eqref{Bp7} is positive respectively negative if $\alpha^2< \alpha_c^2$ respectively $\alpha^2 > \alpha_c^2$,    where
\be
\alpha_c^{-2} :=    \frac{1+ \cn \mu}{1+ \dn \mu} (1 + \dn \mu - \cn \mu).
\label{Bp9}
\ee
(Note that $m \leq \alpha_c^2 \leq 1$.)
Thus the function $\mathcal D(y)$ crosses through zero if the characteristic $\alpha^2$ lies in the interval $(\alpha_c^2, 1)$ --- and remains positive for all $y$ if $\alpha^2$ is in 
the complementary interval $(m, \alpha_c^2)$.

Finally, it remains to consider the situation where  $\sigma=1$ and $\cn \mu <0$.  In this case  the expression in the square brackets in \eqref{Bp1} oscillates between two positive values:
\[
S_- \leq    \sn^2 \mu +  \cn \mu \ts \dn \mu   \ts    y(\xi)  \leq S_+,
\]
where 
\[
S_\pm =\frac{1\mp  \cn \mu}{1+\dn \mu} (1+ \dn \mu \pm \cn \mu)  \geq 0.
\]
As a result, the denominator $\mathcal D$ fills the range $(\mathcal D_\mathrm{min},   \mathcal D_\mathrm{max})$, where
 $\mathcal D_\mathrm{max} \geq  0$ for all $\alpha^2 \in (m,1)$ while $\mathcal D_\mathrm{min}$ is positive respectively negative
for $\alpha^2 < \alpha_c^2$ respectively $\alpha^2 > \alpha_c^2$. This time, the threshold value of $\alpha^2$ is given by
 \be
 \alpha_c^{-2} := \frac{1- \cn \mu}{1+ \dn \mu} (1+ \cn \mu + \dn \mu).
 \label{Bp6}
 \ee
 (Like the transition point defined by \eqref{Bp9}, this critical value satisfies $m \leq \alpha_c^2 \leq 1$.)

 To summarise, in the defocusing situation ($\sigma=-1$)
 equation  \eqref{H2} defines a continuous function of $\xi \in (-\infty, \infty)$ regardless of the value of $\alpha^2$ and sign of $\cn \mu$.
 In the focusing case ($\sigma=1$) 
 the arctangent remains a continuous function of $\xi$ for $\alpha^2 < \alpha_c^2$ but has jump discontinuities when $\alpha^2 > \alpha_c^2$,
 where $\alpha_c^2$ is as in \eqref{Bp9} and \eqref{Bp6} for $\cn \mu>0$ and $\cn \mu<0$, respectively. 
  To obtain a continuous function  $\chi(\xi)$ for  $\alpha^2 > \alpha_c^2$, $\sigma=1$ (and so to retain the relation \eqref{H201}),
  we need to 
consider the analytic continuation of   the function  \eqref{H2} 
to a Riemann surface.

\section{Equivalence of travelling and stationary phase}
\label{AppendixB} 

The construction of the travelling wave solution   in sections \ref{Mod} and \ref{Phase} 
assumes  the agreement of the ``travelling" phase increment  and its stationary counterpart, $\Delta(\xi)$.
Once the phase component of the solution has been determined, 
the procedure needs to be validated by verifying this agreement.  
In this  Appendix, we demonstrate that the difference between $\Delta(\xi)$ and $\chi(\xi)$ (the variable part of the ``travelling" increment) 
 is 
indeed a constant.
Therefore, the travelling and stationary increments can be made equal just by a proper choice of $\kappa$.

Equations \eqref{Y1}, \eqref{Y2} and \eqref{F108} give
  \be
\cot  \Delta (\xi)= 
  \left[ \frac{2 (-\sigma) (1+\sigma A)^2}{\omega I_1} \times 
 \frac{ \cn \mu(\cn \mu+ \dn \mu)}{ \dn^3 \mu}   \mathcal R(\xi)
 \right], 
 \label{H4}
 \ee 
 with $\mathcal R$ as in \eqref{G5}.
The quantity $\cot \Delta$ defines a meromorphic function of $\zeta$,
 with  a pole at the origin:
 \be
\cot \Delta = -        s_{\omega}            \sigma \frac{\sqrt{1+ \sigma A}}{I_1} \mathcal G  \times
 \left( \mathcal R_0 + \frac{1}{\zeta} \right).
\label{Ap18} 
\ee
Here $\zeta$ is as in \eqref{F120}:
\[
\zeta=
 1- m\thinspace \sn^2 \lambda\thinspace  \sn^2(\xi + \lambda),
\]
 while 
  the residue and analytic part of \eqref{Ap18} are straightforward from \eqref{H4} and \eqref{G5}:
\begin{align}  \mathcal G=  2 \frac{ \cn \mu + \dn \mu }{ \dn \mu}, \quad 
\mathcal R_0= - \frac{\sigma A    \ts \cn \mu  \ts  \sn^2 \lambda+  \sn^2 \mu\dn^2\lambda}
{  \sn^2 \mu \ts \dn \mu \ts  (1-m \ts \sn^4 \lambda) }.
\nonumber 
\end{align} 

In the expression for 
 the ``travelling" increment, equation \eqref{H201}, 
the cotangent of $ \chi$ can also be treated as a meromorphic function of $\zeta$. 
To determine poles and residues of that function, we note that $\sn (\xi+\mu)  \ts \sn \ts \xi$ can be written as
 \[
\sn (\xi+\mu)  \ts \sn \ts \xi = \frac{ 1-m \ts \sn^4 \lambda - \zeta}
{m \ts   \sn^2 (\lambda)  \ts \zeta}.
\]
Equation \eqref{H2} gives then
\be
\cot \chi=        s_{\omega}  \ts  \sigma  \frac{\mathcal F}{I_1 \sqrt{1+\sigma A}} 
   \left(
 X_0 + \frac{  X_1 \zeta_1}{\zeta- \zeta_1}  \right), 
\nonumber
\ee
where 
\begin{align}  
\mathcal F=   \frac{\dn \mu}{m \sn^2 \mu} (A-B)^2,    &  \quad     \zeta_1  = 1- m \sn^4 \lambda,  
 \nonumber  \\
 X_1  = -m  \ts  \sn^2 \lambda (1-\alpha^2 \sn^2 \mu), & 
\quad    
 X_0= X_1 - \alpha^2 \cn \mu  \ts \dn \mu.
\nonumber
\end{align}

Making use of the  identity 
 \be
 \cot (\Delta- \chi)= \frac{1+ \cot \chi \cot \Delta}{          \cot \chi   -   \cot \Delta}, 
 \nonumber
 \ee
the trigonometric function  of the difference of the  increments  can be expressed as a ratio
\be 
\label{Ap4}
\cot (\Delta  - \chi ) = \frac{Y_0 +             Y_1  (\zeta - \zeta_1)^{-1}          +             Y_2  { \zeta}^{-1}       }
{Z_0          +      Z_1 (\zeta - \zeta_1)^{-1}                              +        Z_2 \zeta^{-1}         },
\ee
with the  residues $Y_n,  Z_n$ independent of $\xi$. 
Requiring the ratio \eqref{Ap4} to be a constant as a function of  $\xi$ amounts to demanding  it to be a constant as a function of $\zeta$. 
The latter requires the residues of the numerator and denominator  --- at $\zeta_0$, the origin and infinity --- to be proportional:
\be 
\frac{Y_0}{Z_0} = \frac{ Y_1}{ Z_1} = \frac{Y_2}{Z_2}.
\nonumber
\ee
Substituting for $Y_n$ and $Z_n$, the above relations reduce to
\begin{align}
\mathcal F  \, \frac{ X_1   -  X_0  }{1+ \sigma A} +   \mathcal G   \left(\mathcal R_0 + \frac{1}{\zeta_1}\right)= &   0, 
\label{Ap6} \\
\mathcal F^2 X_0 \frac{  X_1 -  X_0 }{1+\sigma A}  +  \mathcal F    \mathcal G  \ts \mathcal R_0  X_1- I_1^2 =  &0.
\label{Ap7}   
\end{align}
Note that every term in \eqref{Ap6} and \eqref{Ap7} is expressible as a rational function of $z = \sigma (A-B)$:
\begin{align}
\mathcal R _0= - \frac{ \ts  \cn \mu    \ts   \dn \mu \ts z+ (\cn \mu+1)(\dn \mu+1) }{2 (\cn \mu + \dn \mu)},    \nonumber     \\
X_0= \frac{m}{z } \left(   m \sn^4 \mu   \frac{1-\cn \mu}{1+ \dn \mu} - \cn \mu  \ts \dn \mu \ts \sn^2 \mu  \right)
-m \left[ \cn \mu  \ts \dn \mu \ts \sn^2 \mu +   \frac{1-\cn \mu}{1+ \dn \mu} (1- m \sn^4 \mu)  \right],
  \nonumber  \\
 X_1 =   \frac{m(1- \cn \mu)}{1+ \dn \mu} \left(  \frac{  m \sn^4 \mu}{z}   + m \sn^4 \mu  -1  \right), \quad
  \frac{ X_1- X_0} {1+ \sigma A} =  \frac{ m \ts \cn \mu \ts \sn^2 \mu}{\dn \mu}  \frac{1 }{z}, \nonumber \\
  I_1^2 = z  \left(  z \dn^2 \mu - m \sn^2 \mu \right)  \times  \left(  \sn^2 \mu - z  \cn^2 \mu\right). 
   \nonumber
\end{align}
Consequently, the left-hand side of \eqref{Ap6} is a polynomial of the first order and that of \eqref{Ap7} second order in $z$.

Verifying that  the coefficient  of each power of $z$ in \eqref{Ap6} and \eqref{Ap7} is zero, we conclude that $  \cot (\Delta -\chi )   $ is indeed $\xi$-independent.
Specifically, 
\be
\cot (\Delta -\chi ) = \frac{Y_2}{Z_2}=         s_{\omega} \sigma    \frac{\mathcal F}{I_1}      \frac{      X_1- X_0                 }{ \sqrt{1+ \sigma A}} .
\label{Ap17}
\ee

\section{Cnoidal wave of the nonlinear Schr\"odinger equations}
\label{cnoNLS} 

Cnoidal wave solutions of the focusing and defocusing nonlinear Schr\"odinger equations
are well known in the literature.
To highlight the similarities with our solution of the Ablowitz–Ladik equation 
and to facilitate the transition to its continuum limit, we briefly reproduce their derivation here.

Substituting an Ansatz $\psi(x, \tau)= p(\eta)e^{i \varphi(\eta) + i \tilde \omega \tau}$ with $\eta=x- \varv \tau$ in equation \eqref{C1},
we decompose it into a system of ordinary differential equations for the real-valued functions $p$ and $\varphi$:
\bea 
\varphi_{\eta \eta} p+ 2 \varphi_\eta p_\eta- \varv  p_\eta=0,    \label{NLS1} \\
p_{\eta \eta} + \varv \varphi_\eta p - \tilde \omega  p -\varphi_\eta^2 p + 2 \sigma p^3=0.  \label{NLS2}
\eea
The order of equation \eqref{NLS1} is reduced by introducing an integrating factor. We obtain
\be
\varphi_\eta = \frac{\varv}{2} +\frac{J_1}{r},  \label{NLS3}
\ee
where $r=p^2$ and $J_1$ is a constant of integration.  Feeding \eqref{NLS3} into \eqref{NLS2}  yields
\be
p_{\eta \eta} + \left( \frac{\varv^2}{4}- \tilde  \omega \right) p + 2 \sigma p^3 - \frac{J_1^2}{p^3}=0.
\label{NLS4}
\ee
An implicit solution of \eqref{NLS4} is given by the quadrature:
\bea
 \pm 2(\eta-\eta_0)=  \int \frac{d r}{\sqrt{-U(r)}},  \label{NLS5} \\
U(r)=   \sigma r^3+\left(\frac{\varv^2}{4}-   \tilde \omega \right) r^2 - J_2r +J_1^2,   \nonumber 
\eea
where $J_2$ and $\eta_0$ are further constants of integration.

The solution $r(\eta)$ is periodic if the cubic polynomial $U(r)$ has three real roots:
\be
U= \sigma (r-a) (r-b) (r-c).  
 \label{NLS7}
\ee
We order the roots as 
\bea
c    \leq    0  \leq    a    \leq   b     &  \quad (\sigma=1);    \label{in1}  \\              
0 \leq   b  \leq   a   \leq c    & \quad (\sigma=-1).    \nonumber       
\eea
Doing the elliptic integral in \eqref{NLS5} and inverting the function $\eta(r)$ gives an explicit expression for $r(\eta)$:
\be
r= a  + (b-a) \cn^2 \left[ \sqrt{\sigma(b-c)} (\eta-\eta_0), m \right],  \quad
m= \frac{b-a}{b-c}.   \label{NLS10} 
\ee

Equations  \eqref{NLS3},  \eqref{NLS5}  and \eqref{NLS7} possess  a scaling invariance: 
\bea
\eta \to \lambda \eta, \quad   r \to \lambda^{-2} r, \quad \varphi \to \varphi;  \nonumber \\
\varv \to \lambda^{-1} \varv, \quad (a,b, c, \tilde \omega) \to \lambda^{-2} (a, b, c, \tilde \omega),
\quad J_1 \to \lambda^{-3} J_1, \quad J_2 \to \lambda^{-4} J_2.
 \label{NLS17} 
 \eea
In what follows, we  fix the value of $b-c$ ---  specifically, 
we
set $b-c=\sigma$ --- while keeping in mind that the family of solutions may be extended at any stage by applying the scaling \eqref{NLS17}.
With this choice, equation \eqref{NLS10} simplifies to
\be
r=a +     \sigma m \cn^2  (\eta-\eta_0, m).   \label{NLS12}
\ee
Here  $m$ ($0 \leq m \leq 1$)   and $a$ ($a \geq 0$)
can be considered independent parameters.
 In the focusing situation ($\sigma=1$)  the inequalities \eqref{in1} imply  $0 \leq a \leq m'$, where $m'=1-m$; 
 in the defocusing case ($\sigma=-1$) we have $a \geq m$. 
The original parameters $\tilde \omega$,         $J_1$,     and  $J_2$ 
are expressible via this new set:
\bea
\tilde \omega= \frac{\varv^2}{4}-m'+m + 3 \sigma a,
 \label{NLS13}
  \\
  J_1=    \tilde s_1 \sqrt{ a(a+\sigma m) (m'-\sigma a)},    \label{NLS21}   \\
J_2=\sigma (mm'-3a^2) +2a (m'-m).   
\nonumber
\eea
In \eqref{NLS21}, $\tilde s_1= \pm 1$ is a sign factor that may be chosen arbitrarily.

Substituting \eqref{NLS12} in \eqref{NLS3} and performing the integration, we recover the phase variable:
\be
\varphi= \frac{\varv}{2}  \eta +  {\tilde s_1}  \sqrt{ \frac{a (m'-\sigma a)}{a+\sigma m}}      \Pi({\tilde \alpha}^2, \mathrm{am} \ts \eta, m),   \label{NLS16} 
\ee
where 
\be
 {\tilde \alpha}^2= \frac{m}{m+\sigma a}
 \label{NLS18}
 \ee
  is positive for $\sigma=1$ and negative for $\sigma=-1$.

The amplitude \eqref{NLS12} and phase \eqref{NLS16} constitute the cnoidal wave solution of 
the focusing and defocusing nonlinear Schr\"odinger equations \eqref{C1}.


    \end{document}